\documentclass[11pt,a4paper]{article}

\usepackage[a4paper,margin=1.03in,includefoot]{geometry}
\usepackage[T1]{fontenc}
\usepackage{lmodern}
\usepackage{iftex}
\ifPDFTeX
  \input{glyphtounicode}
\fi
\usepackage[utf8]{inputenc}
\usepackage[english]{babel}
\usepackage[protrusion=false]{microtype}
\usepackage{amsmath,amssymb,amsthm,mathtools}
\usepackage{booktabs}
\usepackage{tabularx}
\usepackage[section]{placeins}
\usepackage{graphicx}
\usepackage{subcaption}
\usepackage{float}
\usepackage{xcolor}
\usepackage{tikz}
\usetikzlibrary{arrows.meta,positioning}
\usepackage[numbers,sort&compress]{natbib}
\usepackage{xurl}
\usepackage[hidelinks]{hyperref}
\usepackage[nameinlink,noabbrev]{cleveref}
\newcommand*{\doi}[1]{doi:~\href{https://doi.org/#1}{\nolinkurl{#1}}}

\theoremstyle{plain}

\theoremstyle{definition}

\theoremstyle{remark}

\newcommand{\IQM}{IQM Emerald}
\newcommand{\Rrec}{\widehat R_{\mathrm{rec}}}
\newcommand{\Rctrl}{\widehat R_{\mathrm{ctrl}}}
\newcommand{\protocol}[1]{\textsc{#1}}

\title{From Round-Trip State Echo to Error Recovery:\\
Snapshot-Resolved Quantum-Hardware Diagnostics}
\author{Isaac Barouch Essayag\\
\small MIGAL--Galilee Research Institute/Tel-Hai University of Kiryat Shmona\\
\small in the Galilee, P.O. Box 831, Kiryat Shmona 1101602, Israel
\and
Aryeh Lev Zabokritskiy (Yohananov)\\
\small Department of Computer Science, Tel-Hai University of Kiryat Shmona\\
\small in the Galilee, Israel; MIGAL--Galilee Research Institute,\\
\small P.O. Box 831, Kiryat Shmona 1101602, Israel}
\date{}
\hypersetup{
  pdftitle={From Round-Trip State Echo to Error Recovery: Snapshot-Resolved Quantum-Hardware Diagnostics},
  pdfauthor={Isaac Barouch Essayag; Aryeh Lev Zabokritskiy (Yohananov)},
  pdfsubject={Repeated hardware tests of quantum communication and single-deletion recovery},
  pdfkeywords={quantum benchmarking, state transfer, quantum deletion codes, superconducting qubits, trapped ions},
  pdflang={en-US}
}

\begin{document}
\maketitle

\begingroup
\microtypesetup{expansion=false}
\begin{abstract}
We examine which conclusions from quantum-hardware benchmarks persist when
the same programs are run again. On eight fixed routes of IQM Emerald,
round-trip state echo (RTSE), a diagnostic of state return after a round trip,
gave mean scores near $0.84$ that differed by only $0.00125$ between two
executions. Individual route estimates nevertheless changed by up to $0.06$
in both directions. The mean difference between RTSE and measurement-free teleportation along
the same routes shrank from $0.046$ to $0.014$ and was no longer statistically
resolved on repetition; none of the four prespecified communication targets
met its replication rule. Recovery with an existing four-qubit
single-deletion code also changed on repetition: after discarding one encoded
qubit in the circuit model, the IQM mean fell from $0.738$ to $0.624$ and
failed the replication test against the two-thirds operational reference.
In a separate IonQ experiment, recovery means of $0.91$ and $0.92$ passed
tests against that reference and against a deliberately incorrect decoder
in both executions. These results concern the tested workloads and
executions. They show why a nearly unchanged aggregate benchmark score needs
to be interpreted alongside individual routes, direct task comparisons, and
repeated executions.
\end{abstract}
\endgroup


\section{Introduction}
\label{sec:introduction}

A quantum benchmark records how well a processor performs a specified task.
Component benchmarks characterize individual operations, whereas quantum
volume and volumetric benchmarks describe capabilities over larger circuits
\citep{cross2019quantumvolume,blumekohout2020volumetric}. Full-stack and
application-oriented studies show why performance depends on the workload
being assessed \citep{proctor2022capabilities,proctor2025benchmarking,lubinski2023application,hines2024fullstack}.
Using a benchmark to choose a circuit or a qubit placement also requires
understanding how well its conclusions persist over time. Calibration-aware
mapping studies address variation across placements and days, while QKNOB
isolates routing overhead in circuit transformation
\citep{murali2019noiseadaptive,li2025qknob}. Dasgupta and Humble bounded
departures from ideal single-qubit outcomes under fluctuating device noise
\citep{dasgupta2022reproducibility}. Here we ask whether conclusions drawn
from fixed programs survive repetition under decision rules set before the
new outcomes are known.

Communication protocols provide task-specific tests of quantum hardware.
Huang et al.\ benchmarked state transfer against a classical strategy
\citep{huang2021statetransfer}; Zhukov et al.\ demonstrated superdense coding
and BB84 on IBM processors in a study published online in 2018 and in the
2019 journal volume \citep{zhukov2019protocolbenchmarks}.
Meirom et al.\ developed a seven-protocol framework in 2025 that includes the
do-nothing, Bell-state transfer, teleportation, and entanglement-swapping
protocols, and reported earlier teleportation and entanglement-swapping
experiments by Tal Mor, Chen Mechel, and Rotem Liss around 2018
\citep{meirom2025protocols}. Mayo, Mor, and Weinstein subsequently applied
that framework to superconducting processors and comparisons with
trapped-ion hardware \citep{mayo2026ibm,mayo2026crossplatform}. A separate
teleportation benchmark studied IBM and Rigetti hardware
\citep{marquez2025teleportation}. These task-based benchmarks provide the
setting for our comparisons of repeated communication outcomes.

Our first diagnostic is \emph{round-trip state echo} (RTSE), a variant of
the do-nothing protocol. Its score measures how well a qubit state returns
to its initial location after a round trip. A related study coauthored by
one of the present authors examined how the predecessor's return score
changes with swap distance on IBM and Rigetti processors
\citep{mayo2026howfar}. We compare RTSE with that predecessor on two IQM
routes, with measurement-free teleportation along eight fixed routes, and
separately compare Bell-state transfer with entanglement swapping. The
comparisons distinguish the round-trip diagnostic from direct communication
outcomes and compare two entanglement workloads on the same hardware.
Mirror-circuit and Loschmidt-echo studies provide related diagnostics based
on reversibility \citep{peters2022timereversal,proctor2022mirror}.
Throckmorton and Das Sarma studied round-trip SWAP fidelity in noisy spin
chains \citep{throckmorton2020swap}, and QPing studies return-based
diagnostics in quantum networks \citep{miguelramiro2026qping}. Layer-fidelity
studies characterize long chains through a complementary benchmark
\citep{mckay2023layerfidelity,lozano2026layerfidelity}. Our emphasis is on
what repeated protocol outcomes reveal about fixed routes and workloads.

Deletion recovery provides a further task whose performance can be assessed
on repetition. Hagiwara and Nakayama established a single-deletion code of
optimal length four \citep{hagiwara2020deletion}. An earlier hardware
implementation tested one deletion position and three input states
\citep{nakayama2020singledeletion}. Our experiments extend that coverage to
all four deletion positions and four input states, with repeated executions
on IQM and, separately, IonQ. The tested task is recovery after one encoded
qubit is discarded in the circuit model.

On eight fixed routes of IQM Emerald, two executions of identical programs
gave aggregate RTSE scores near $0.84$ differing by only $0.00125$.
Individual route estimates moved by up to $0.06$ in both directions, partly
canceling in the average. The aggregate difference between RTSE and
teleportation shrank from $0.046$ to $0.014$ and was no longer statistically
resolved in the second execution. The largest route-level difference in the
first execution shrank from $0.211$ to $0.045$ on repetition. The case of
false reassurance, in which that route passed the RTSE test but failed the
teleportation test, also did not meet its replication rule.

The comparison between entanglement swapping and Bell-state transfer behaved
differently: on their four matched routes, the estimated difference changed
little, from $0.043$ to $0.041$. Its uncertainty interval nevertheless
included zero on repetition. None of the four communication targets fixed
after the first execution met its replication rule. The nearly unchanged
swapping--Bell difference illustrates the distinction between an estimated
effect and a statistically resolved one. In the supporting length
study on two IQM routes, both RTSE and its do-nothing predecessor scored
lower at 10 edges than at 2 edges in each of two executions. The
prespecified comparison did not confirm a smaller loss for RTSE;
descriptively, its score fell more.

The recovery experiments gave different replication outcomes on the two
devices. On IQM, the mean recovery score fell from $0.738$ to $0.624$ on a
repeat with a different submitted order, declined at all four fixed
placements, and failed its prespecified replication test against the
two-thirds operational reference. In the separate IonQ experiment,
recovery means of $0.91$ and $0.92$ exceeded that reference and outperformed
a deliberately incorrect decoder in both executions. These results characterize each
device under its own programs and controls.
Exceeding two-thirds is an operational benchmark result here, not a
certification that the hardware outperforms every classical strategy: the
classical state-estimation bound requires additional assumptions about
preparation, readout, and the tested channel \citep{massar1995optimal}.

Together, the results describe complete workloads on the tested routes and
placements, with two executions per experiment. They support reading
aggregate benchmarks alongside route-level outcomes, direct task
comparisons, and repetitions. A high round-trip score still leaves open
whether the state reached the far endpoint within that circuit. Testing
whether RTSE can predict performance on another task remains a separate
question in the broader assessment of quantum-computing capability
\citep{hothem2024capability}.

Section~\ref{sec:profile} defines the tasks and measured outcomes, and
Section~\ref{sec:design} describes the experimental setting.
Sections~\ref{sec:communication} and~\ref{sec:length-experiment} present the
communication and route-length results on IQM;
Sections~\ref{sec:iqm-recovery} and~\ref{sec:ionq-experiment} present the IQM
and IonQ recovery results. Section~\ref{sec:discussion} discusses their
implications and limitations, and Section~\ref{sec:conclusion} concludes.

\section{Tasks and measured outcomes}
\label{sec:profile}

\begin{figure}[htbp]
  \centering
  \includegraphics[width=\textwidth,
    alt={Five workload flows: round-trip state echo, measurement-free teleportation, Bell-state transfer, entanglement swapping, and deletion recovery. The communication panels give per-program resources.}]{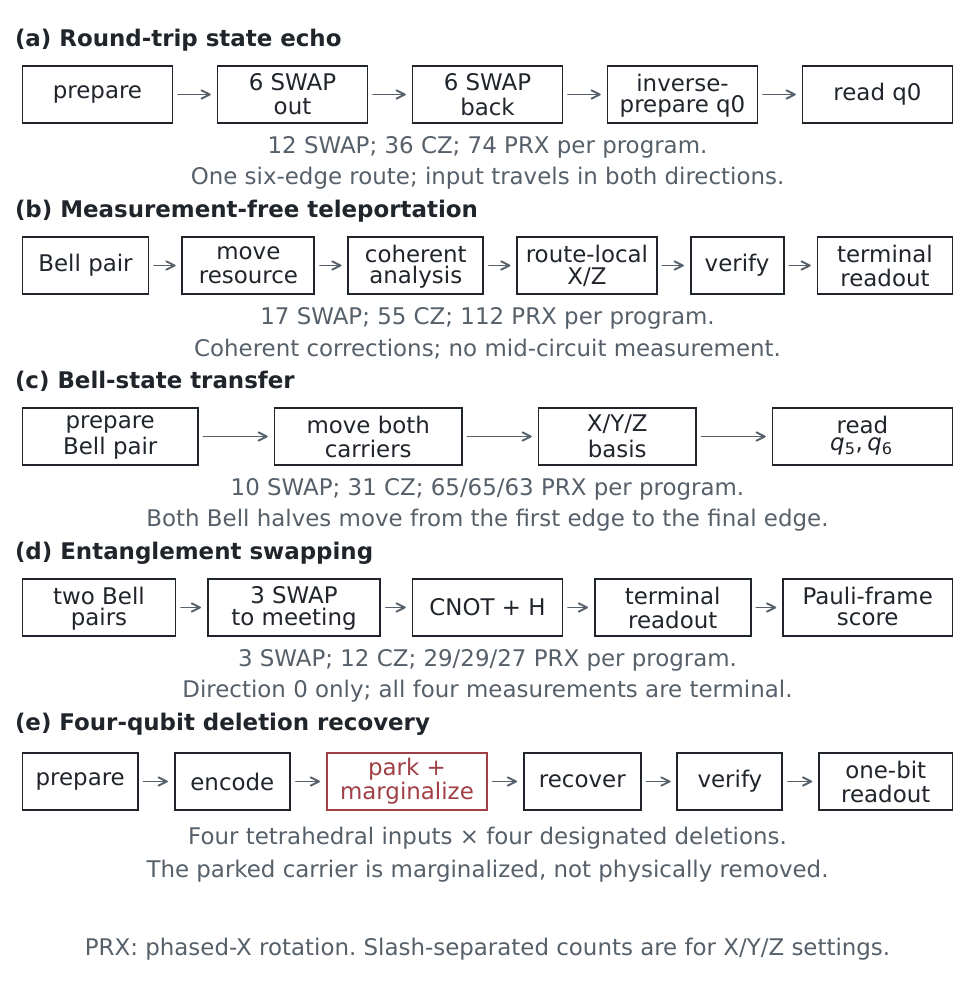}
  \caption{Five workloads with different state transformations and readouts.
  Route positions run from $q_0$ at the initial qubit to $q_6$. Each communication
  panel gives the logical stages and resource counts for
  one submitted program; slash-separated PRX counts follow the $X/Y/Z$ basis
  settings. Direction zero identifies the route chosen in advance for
  entanglement swapping from each initial qubit's two routes. Recovery combines encoding,
  discard in the circuit model (red), coherent decoding,
  and a return test; its separately compiled resources appear in
  \cref{tab:qec-resources}. Arrows indicate the
  order of operations.}
  \label{fig:profile}
\end{figure}

Each task specifies an input ensemble, a complete circuit, a physical placement
or virtual-qubit assignment, a measured statistic, and an operational reference
value.  RTSE provides the initial diagnostic, followed by communication tasks
on the same routes and a separate recovery test (\cref{fig:profile}).  Each of
the two communication comparisons pairs workloads on the same
directed physical route and in the same execution, using corresponding input or
measurement settings: RTSE with measurement-free teleportation, and
entanglement swapping with Bell-state transfer.  Their
gate counts, depth, duration, and native resources may differ; each comparison
therefore concerns the complete workloads.

RTSE measures the combined performance of preparation, the compiled round
trip, and readout on the chosen route.  Locating a failure or assessing a
communication or recovery task requires a separate test.  We submitted all communication families regardless of the RTSE outcomes and compared each
route's RTSE result with the measurement-free teleportation result on that route; \cref{sec:communication} also tests whether
a combination of an RTSE pass and a teleportation failure observed on one route recurs on that route.

For the compiled-resource annotations below, our convention for the native
phased-$X$ rotation is
\[
  \operatorname{PRX}(\theta,\phi)
  \coloneqq R_Z(\phi)R_X(\theta)R_Z(-\phi).
\]
Every displayed CZ, PRX, and algorithmic-SWAP count is per submitted program.  A slash-separated PRX triple is
ordered as $X/Y/Z$ for the three separately compiled Pauli-basis programs.

\subsection{A common single-qubit input ensemble}

RTSE, its do-nothing predecessor, teleportation, and recovery are evaluated on four fixed pure
states whose Bloch vectors are
\begin{equation}
  \mathbf v_0\coloneqq\frac{(1,1,1)}{\sqrt 3},\quad
  \mathbf v_1\coloneqq\frac{(1,-1,-1)}{\sqrt 3},\quad
  \mathbf v_2\coloneqq\frac{(-1,1,-1)}{\sqrt 3},\quad
  \mathbf v_3\coloneqq\frac{(-1,-1,1)}{\sqrt 3}.
  \label{eq:tetrahedron}
\end{equation}
Let $\lvert\psi_s\rangle$ denote the pure state with Bloch vector
$\mathbf v_s\eqqcolon(v_{s,x},v_{s,y},v_{s,z})$, and set
$\rho_s\coloneqq\lvert\psi_s\rangle\!\langle\psi_s\rvert$.  These states satisfy
\begin{equation}
  \frac14\sum_{s=0}^{3}\mathbf v_s=0,\qquad
  \frac14\sum_{s=0}^{3}\mathbf v_s\mathbf v_s^{\mathsf T}=\frac{I_3}{3};
  \quad
  \frac14\sum_{s=0}^{3}\rho_s^{\otimes2}=\frac{\Pi_{\mathrm{sym}}}{3},
  \label{eq:tetrahedron-two-design}
\end{equation}
where $I_3$ is the $3\times3$ identity and $\Pi_{\mathrm{sym}}$ projects onto
the two-qubit symmetric subspace.
Thus the four projectors form a qubit complex-projective two-design
\citep{renes2004sic,scott2006tight}.

The original do-nothing construction prepares a randomly chosen
single-qubit state, fixed within each experiment. Here, for every workload that uses these inputs, including our implementation of the predecessor, we fix the compiled convention
\begin{equation}
  \theta_s\coloneqq\arccos(v_{s,z}),\qquad
  \phi_s\coloneqq\operatorname{atan2}(v_{s,y},v_{s,x}),\qquad
  U_s\coloneqq R_Z(\phi_s)R_Y(\theta_s),
  \label{eq:input-unitaries}
\end{equation}
so that $U_s\lvert0\rangle$ equals $\lvert\psi_s\rangle$ up to a global
phase.  The preparation applies $R_Y(\theta_s)$ followed by
$R_Z(\phi_s)$ (on IQM the $R_Z$ rotation is absorbed into the phases of later
\texttt{PRX} gates), and inverse preparation applies the inverses of these
gates in reverse order.

After the
complete placed circuit, $U_s^\dagger$ is applied at the designated output (the predecessor instead applies it remotely, before the return leg) and
that output is measured in the computational basis.  Let $\widehat p_s(0)$ be
the observed proportion of outcome zero for input setting $s$, an estimate of the corresponding probability $p_s(0)$.  We call
\begin{equation}
  \widehat R\coloneqq\frac14\sum_{s=0}^{3}\widehat p_s(0)
  \label{eq:return-score}
\end{equation}
the end-to-end return score; ``return'' refers to the prepared input returning to $\lvert0\rangle$ under inverse preparation, so the score also applies to one-way teleportation and to recovery.  With ideal preparation, inverse preparation,
and measurement, $p_s(0)$ is the target-state overlap.  The equally weighted mean of these return overlaps
equals the Haar pure-state average when one fixed qubit channel acts for
all $s$ and preparation and measurement are ideal.  The hardware implementation
compiles each state setting separately. Its reported quantity is the
equal-weight average of four fixed cell proportions, one per separately compiled program, allowing preparation,
compilation, and readout to depend on the input setting.  The hardware return score includes uncorrected
state preparation and measurement (SPAM) errors.

The value $2/3$ comes from optimal single-copy qubit-state estimation
\citep{massar1995optimal}: it bounds the tetrahedral average for a fixed
measure-and-prepare, or entanglement-breaking, channel that receives one
unknown qubit without its classical state label, assuming ideal trusted
preparation and readout.  Our separately compiled hardware cells include
input-dependent preparation, compilation, and readout.  These cells therefore do not meet the bound's assumptions, and we use
$2/3$ only as a prospectively chosen operational reference for these complete
workloads: exceeding it is a benchmark criterion, not a certified separation from measure-and-prepare channels. For recovery, the criterion concerns the score averaged over the four deletion positions. Establishing recovery at every position, fault tolerance, or a benefit from physical error correction requires further evidence.

\subsection{Round-trip state echo and its predecessor}

\begin{figure}[htbp]
  \centering
  \includegraphics[width=\textwidth,
    alt={Two twelve-swap round trips differ in inverse-preparation placement and terminal checks: the predecessor's full specification checks seven qubits; RTSE checks only the initial qubit.}]{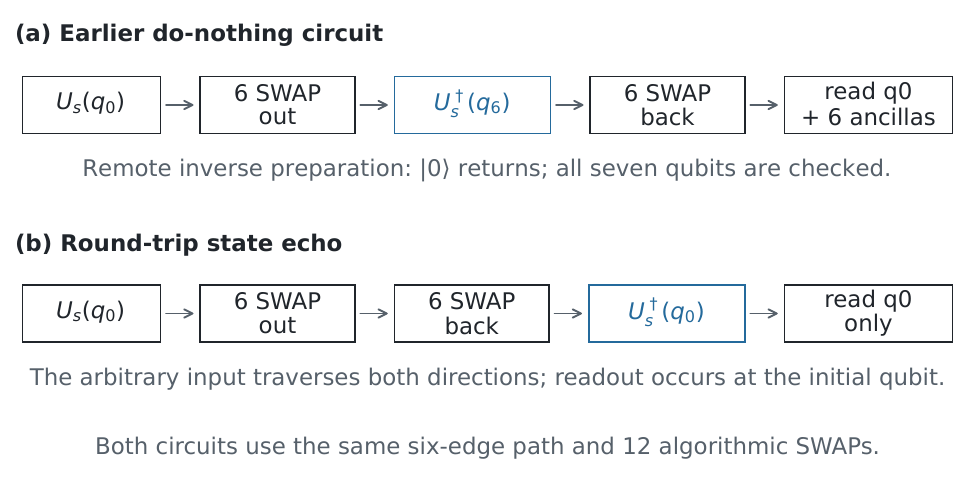}
  \caption{RTSE moves inverse preparation to the initial qubit and reduces terminal
  readout to one qubit. Both circuits use a 12-swap round trip on the same
  six-edge route. The predecessor applies $U_s^\dagger$ at the endpoint,
  returns $\lvert0\rangle$, and in its full specification reads the initial qubit and six route ancillas.
  RTSE returns $\lvert\psi_s\rangle$ before applying $U_s^\dagger$ at the
  initial qubit. Blue highlights the inverse-preparation stage. Both scores test return
  at the initial qubit; endpoint signal is tested separately by one-way circuits.}
  \label{fig:rtse-comparison}
\end{figure}

For the \protocol{round-trip state echo} (\protocol{RTSE}), an
input state is prepared at the initial qubit of a simple physical route, moved to the
far endpoint by a sequence of swaps, returned along the reversed route,
inverse-prepared at the same initial qubit, and measured; its score is the tetrahedral mean in \cref{eq:return-score}.  This is a variant of the
do-nothing protocol of Meirom et~al.~\citep{meirom2025protocols}.  Both variants prepare and read the work qubit at
the initial qubit and, on a six-edge route, execute a 12-swap round trip.  The
predecessor, in its full specification, additionally measures all six returned route ancillas and applies
its operational reference to each; RTSE records only the output at the initial qubit and
deliberately drops those ancilla-return checks, which the predecessor defines but usually omits in its reported experiments.  The predecessor applies
$U_s^\dagger$ remotely before the return, so the arbitrary state
$\lvert\psi_s\rangle$ traverses the outward leg but the reference state
$\lvert0\rangle$ traverses the return leg.  RTSE delays $U_s^\dagger$ until
after the return: the same arbitrary input traverses both directions, while
$U_s$, $U_s^\dagger$, and readout remain at the initial qubit, and the far endpoint
requires no state-dependent inverse gate.  RTSE therefore needs less control at the far endpoint
and fewer terminal readouts than the predecessor's full specification, but it requires the arbitrary state to remain
coherent on both legs; both variants use the same 12-swap round trip (\cref{fig:rtse-comparison}). The variants
share the ideal target at the initial qubit but carry different states on the return leg
and read different outputs.

A round-trip echo can also score highly for a process that never transports the
state, for example one whose operations cancel. The length study
(\cref{sec:length-experiment}) therefore tests whether the RTSE score falls as the
route is extended along fixed shortest paths, and adds one-way circuits that invert
the preparation at the far endpoint to test for input-dependent signal there;
whether the state arrives at the far endpoint in the RTSE circuit itself remains unverified.

For the predecessor, the length study records two distinct outcomes.  Its
primary score is the marginal at the initial qubit, so it can be compared with RTSE at the same
readout location.  A secondary joint score requires the initial qubit and every
returned route ancilla to be zero.  This joint all-zero probability is our
aggregation of the predecessor's individually measured outputs; the cited
protocol applies its reference separately to the work qubit and to each
ancilla.  The joint statistic preserves that wider terminal readout.  Route
length $L$ is the one-way swap distance; a positive-length round trip therefore
contains $2L$ algorithmic swaps.

Layer-fidelity methods select and monitor long chains through a metric
derived from randomized benchmarking
\citep{mckay2023layerfidelity,lozano2026layerfidelity}. Here we instead hold
the routes and complete programs fixed and measure their protocol outcomes.

RTSE belongs to the broader family of mirror-circuit and Loschmidt-echo
diagnostics, which use reversibility as an inexpensive proxy for an executed
process \citep{peters2022timereversal,proctor2022mirror}.  It is also related
to the path-based active ``bouncing'' strategy of QPing, in which one half of a
Bell pair traverses a network path outward and back before a Bell measurement at the origin and a thresholded fidelity decision
\citep{miguelramiro2026qping}.  QPing assumes entanglement-distribution
functionality in a quantum network; RTSE instead executes a separable
tetrahedral probe as a SWAP-chain circuit within one processor and scores it
by inverse preparation and $Z$ readout.  The present study
implements this route-resolved probe and evaluates it on hardware at fixed
lengths.

The relation to QPing's Bell-pair test is exact for a fixed channel and
ideal SPAM.  Under these assumptions, let
$\mathcal E_{\mathrm{rt}}$ be one fixed effective qubit channel for the complete
outward-and-return path.  Then the two-design identity gives, for the ideal RTSE score,
\begin{equation}
  R_{\mathrm{RTSE}}^{\mathrm{ideal}}
  \coloneqq\frac14\sum_{s=0}^{3}\operatorname{Tr}
    [\rho_s\mathcal E_{\mathrm{rt}}(\rho_s)]
  =F_{\mathrm{avg}}(\mathcal E_{\mathrm{rt}})
  =\frac{2F_e(\mathcal E_{\mathrm{rt}})+1}{3},
  \label{eq:rtse-entanglement-fidelity}
\end{equation}
where $F_e$ is the Bell-pair entanglement fidelity
\citep{horodecki1999teleportation,nielsen2002average}.  QPing's active bounce estimates the corresponding
Bell-overlap functional but uses its own task- and time-dependent network
threshold.  Thus, for the same fixed channel and ideal preparation and readout,
$R_{\mathrm{RTSE}}^{\mathrm{ideal}}>2/3$ if and only if $F_e>1/2$.

\subsection{Communication comparisons on the same routes}

Let $\mathcal R\coloneqq\{15,19,35,39\}$ be the identifiers of the four fixed initial qubits.  For
$r\in\mathcal R$ and direction label $j\in\{0,1\}$, write the ordered six-edge
communication route as
\begin{equation}
  \pi_{r,j}\coloneqq(q_{r,j,0},\ldots,q_{r,j,6}),\qquad q_{r,j,0}=r.
  \label{eq:route-notation}
\end{equation}
The order fixes the orientation from the initial qubit to the endpoint; the nominal direction label $j$ distinguishes the two fixed routes from that initial qubit.  The eight sequences are listed in
\cref{tab:routes}.

For the circuit descriptions below, abbreviate $q_{r,j,k}$ to $q_k$ on the
chosen route.  In \protocol{Bell-state transfer}, a Bell pair is created on
$(q_0,q_1)$.  Five successive swaps move the half at $q_1$ to $q_6$;
five more move the half at $q_0$ to $q_5$.  Both halves therefore arrive
at the final route edge $(q_5,q_6)$.  Their target overlap with
$\lvert\Phi^+\rangle\coloneqq(\lvert00\rangle+\lvert11\rangle)/\sqrt2$ is estimated
in the three Pauli bases.  For $B\in\{X,Y,Z\}$, let
$e^{(B)}_{5,t},e^{(B)}_{6,t}\in\{0,1\}$ be the outcomes at $q_5,q_6$ on shot
$t$, where zero and one represent the $+1$ and $-1$ eigenvalues after the
submitted basis rotations.  For the $n_B$ shots in that basis, define
\begin{equation}
  \widehat{\langle BB\rangle}
  \coloneqq\frac{1}{n_B}\sum_{t=1}^{n_B}
    (-1)^{e^{(B)}_{5,t}+e^{(B)}_{6,t}}.
  \label{eq:bell-correlator}
\end{equation}
The Bell-projector identity then gives the estimator
\begin{equation}
  \widehat F_{\Phi^+}
  \coloneqq\frac{1+\widehat{\langle XX\rangle}
            -\widehat{\langle YY\rangle}
            +\widehat{\langle ZZ\rangle}}{4}.
  \label{eq:bell-fidelity}
\end{equation}
For one common state measured with trusted Pauli measurements, the maximum
overlap of a separable state with $\lvert\Phi^+\rangle$ is $1/2$, the square
of its largest Schmidt coefficient.  Here the three bases are separately
compiled end-to-end programs, and the raw correlators include preparation and
readout.  Thus $\widehat F_{\Phi^+}$ is an uncorrected end-to-end Bell-overlap
score with operational reference $1/2$, used in the same sense as $2/3$.

The same estimator and reference are used after
\protocol{entanglement swapping}, with terminal outcomes at the intermediate
station incorporated as a Pauli-frame sign.  Here Bell pairs are prepared on
$(q_0,q_1)$ and $(q_5,q_6)$.  Three swaps move the half at $q_1$ through
$q_2,q_3$ to $q_4$, where a CNOT from $q_4$ to $q_5$ followed by $H$ on
$q_4$ performs the Bell analysis.  Terminal readout is ordered as
$(q_0,q_4,q_5,q_6)$: $m_{4,t},m_{5,t}$ are the outcomes at $q_4,q_5$,
and $e_{0,t},e_{6,t}$ are the route-endpoint outcomes at $q_0,q_6$.
For each $B\in\{X,Y,Z\}$, all four unsuperscripted bits in
$c_{B,t}$ below come from that basis-$B$ program, and $t$ is local to that
program.  The per-shot corrected Pauli products are
\begin{equation}
  c_{X,t}\coloneqq(-1)^{e_{0,t}+e_{6,t}+m_{4,t}},\quad
  c_{Y,t}\coloneqq(-1)^{e_{0,t}+e_{6,t}+m_{4,t}+m_{5,t}},\quad
  c_{Z,t}\coloneqq(-1)^{e_{0,t}+e_{6,t}+m_{5,t}},
  \label{eq:swapping-pauli-frame}
\end{equation}
where exponents are evaluated modulo two and
$\widehat{\langle BB\rangle}_{\mathrm{swap}}\coloneqq n_B^{-1}\sum_t c_{B,t}$.
With this Bell-analysis bit ordering, outcome $(m_4,m_5)$ gives the endpoint
byproduct $(I\otimes X^{m_5}Z^{m_4})\lvert\Phi^+\rangle$.  Because
$\lvert\Phi^+\rangle$ has $XX,YY,ZZ$ expectations $(+1,-1,+1)$, conjugating
by the byproduct produces exactly the three sign corrections in
\cref{eq:swapping-pauli-frame}.  The sample means $\widehat{\langle BB\rangle}_{\mathrm{swap}}$ are inserted into
\cref{eq:bell-fidelity} to define $\widehat F_{\mathrm{swap}}$.  This is the
circuit-specific Pauli-frame form of standard entanglement swapping
\citep{zukowski1993eventready} and a deferred-measurement
variant of the usual measurement-and-feed-forward protocol: it preserves the
ideal Bell-state target while avoiding mid-circuit measurement as an extra
hardware requirement.

In \protocol{measurement-free teleportation}, let $a$ carry
$\lvert\psi_s\rangle$ and let $b,c$ carry $\lvert\Phi^+\rangle$.  At the ideal
logical level,
\begin{equation}
  \mathrm{CZ}_{a,c}\,\mathrm{CNOT}_{b\to c}\,H_a\,
  \mathrm{CNOT}_{a\to b}
  \bigl(\lvert\psi_s\rangle_a\lvert\Phi^+\rangle_{bc}\bigr)
  =\lvert+\rangle_a\lvert+\rangle_b\lvert\psi_s\rangle_c.
  \label{eq:coherent-teleportation-identity}
\end{equation}
Thus coherent controlled corrections replace Bell-basis measurement,
classical feed-forward, and conditioned corrections in the teleportation
protocol \citep{bennett1993teleportation}, while preserving the ideal
destination state; the same measurement-free form was used as a
benchmark on a single processor by \citet{marquez2025teleportation}, who used
inverse preparation and payloads of random gates on IBM and Rigetti hardware.  On route
$\pi_{r,j}=(q_0,\ldots,q_6)$, the submitted circuit prepares the input on $q_0$
and the Bell pair on $q_1,q_2$, then transports $q_2$ to $q_6$.  After
$\mathrm{CNOT}_{q_0\to q_1}$ and $H_{q_0}$, it moves the $q_1$ control to $q_5$,
applies $\mathrm{CNOT}_{q_5\to q_6}$, and returns that control.  It then moves the
$q_0$ control to $q_5$ and applies $\mathrm{CZ}_{q_5,q_6}$.  Only $q_6$ is
inverse-prepared and measured, and its score is \cref{eq:return-score}.  This
route-local realization has the same ideal teleportation target, but its
hardware score characterizes this particular circuit.  For each directed route
we define
the paired RTSE--teleportation contrast
\begin{equation}
  \widehat\Delta_{\mathrm{tel}}
  \coloneqq\widehat R_{\mathrm{RTSE}}-
   \widehat R_{\mathrm{teleportation}}.
  \label{eq:teleportation-penalty}
\end{equation}
Here $\widehat R_{\mathrm{RTSE}}$ and $\widehat R_{\mathrm{teleportation}}$ are the
return scores in \cref{eq:return-score} of the two workloads on the same route and in
the same execution.  The RTSE--teleportation contrast asks whether measurement-free
teleportation achieves a lower return score than RTSE on that route; a positive
value means that RTSE scored higher. Circuit-resource and observable differences are part of
this workload contrast; a causal estimate of teleportation cost would require
controls that separate their contributions.

Entanglement swapping was executed on one prespecified route from each initial
qubit, the route with $j=0$; we call this subset direction zero, as in
\cref{tab:routes}.  Within each execution, we compare it with Bell-state transfer on those same four directed routes.
Writing $\widehat F_{\mathrm{Bell},r,j}$ for the Bell-state-transfer score $\widehat F_{\Phi^+}$ of
\cref{eq:bell-fidelity} on route $\pi_{r,j}$, and $\widehat F_{\mathrm{swap},r,0}$ for $\widehat F_{\mathrm{swap}}$
on route $\pi_{r,0}$, define
\begin{equation}
  \overline\Delta_{\mathrm{swap}}
  \coloneqq\frac14\sum_{r\in\mathcal R}
   \left(\widehat F_{\mathrm{swap},r,0}
        -\widehat F_{\mathrm{Bell},r,0}\right).
  \label{eq:swapping-bell-contrast}
\end{equation}
The swapping--Bell contrast compares the Bell-overlap scores attained by two
entanglement workloads on the same four direction-zero routes.
The separately reported Bell-family mean averages all eight directed Bell
routes, while \cref{eq:swapping-bell-contrast} uses the four-route
direction-zero subset for both terms. The output pairs also differ: swapping
reads $(q_0,q_6)$, whereas Bell-state transfer reads $(q_5,q_6)$. A positive value
means that swapping attained the higher Bell-overlap score on those
same routes, with each workload read at its own output pair.

\subsection{Recovery after discarding one qubit of the codeword}

Recovery tests the processor's execution of an encoding and decoding circuit:
the input must be reconstructed from the surviving carriers after one
subsystem is discarded. On IQM, the experiment uses the first five qubits of each
direction-zero communication route and measures the performance and repeatability of this further
workload directly.

\begin{figure}[htbp]
  \centering
  \includegraphics[width=\textwidth,
    alt={Exact logical encoder and coherent decoder, read left to right. The decoder acts on the three survivors in inherited order and a zero-initialized ancilla; the text lists every gate.}]{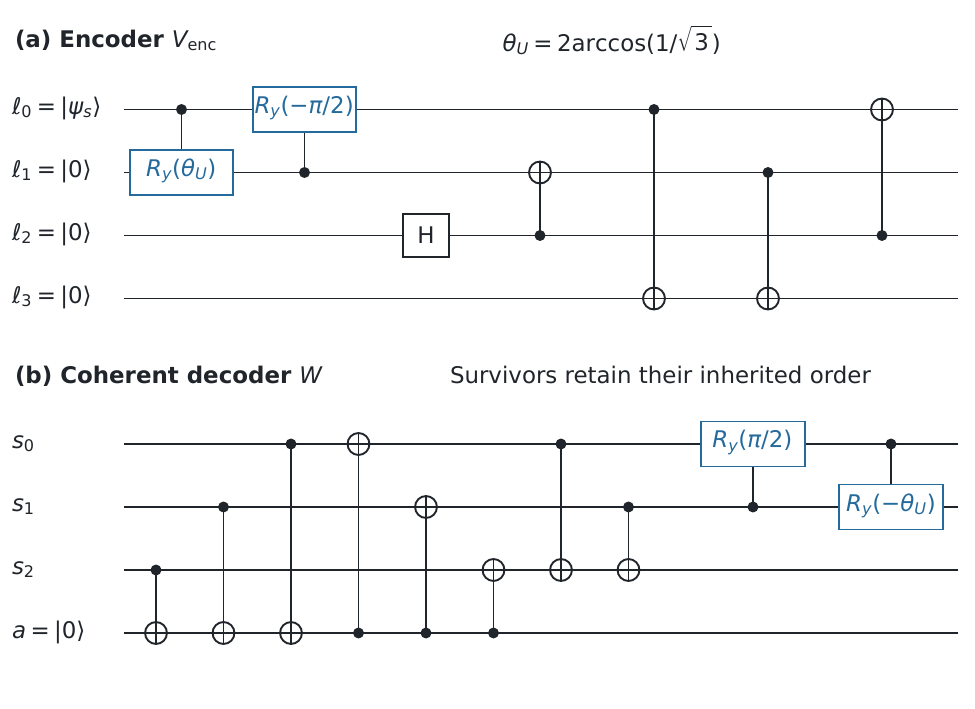}
  \caption{Logical circuits for the four-qubit deletion code
  \citep{hagiwara2020deletion}. Panel (a) implements $V_{\mathrm{enc}}$ on
  input $\lvert\psi_s\rangle$ at $\ell_0$ and three zero-initialized wires.
  Panel (b) implements $W$ on the survivors $(s_0,s_1,s_2)$ in inherited
  order and an ancilla initialized to $\lvert0\rangle$. Blue boxes are
  $R_y$ targets; a joined filled dot denotes a control, and a circled plus
  is a CNOT target. The same decoder acts for every deletion position. These logical blocks are surrounded by preparation, parking and
  marginalization, routing, native lowering, inverse preparation, and
  readout as specified in the text.}
  \label{fig:qec-circuit}
\end{figure}

We use the four-qubit code of Hagiwara and Nakayama
\citep{hagiwara2020deletion}.  Its logical codewords are
\begin{equation}
 \lvert 0_L\rangle\coloneqq\frac{\lvert0000\rangle+\lvert1111\rangle}{\sqrt2},
 \qquad
 \lvert 1_L\rangle\coloneqq\frac{1}{\sqrt6}
 \sum_{\substack{x\in\{0,1\}^4\\\mathrm{wt}(x)=2}}\lvert x\rangle.
 \label{eq:deletion-codewords}
\end{equation}
Each single-carrier reduced state is $I_2/2$, where $I_2$ is the one-qubit
identity, independently of the logical input: the two codeword marginals
are both $I_2/2$, and the single-carrier reduction of
$\lvert0_L\rangle\!\langle1_L\rvert$ vanishes. Thus the discarded carrier
alone contains no information about the input; the decoder recovers that
information from correlations among the survivors.
We use the encoding isometry and the coherent measurement-free decoder
shown in Figures~1 and~3 of the arXiv version of \citet{hagiwara2020deletion}
(arXiv:2001.08405v1), redrawn in \cref{fig:qec-circuit} in controlled-$R_y$
notation.
We add the preparation,
compilation, discard, and readout operations needed to evaluate the complete
workload on hardware.

For a designated position $d\in\{0,1,2,3\}$, the ideal deletion channel is the partial trace
\begin{equation}
  \mathcal D_d(\rho)\coloneqq\operatorname{Tr}_d(\rho).
  \label{eq:synthetic-deletion}
\end{equation}
Let $V_{\mathrm{enc}}:\mathbb C^2\rightarrow(\mathbb C^2)^{\otimes4}$ be the
encoding isometry $V_{\mathrm{enc}}\lvert b\rangle\coloneqq\lvert b_L\rangle$.  After
deleting carrier $d$,
write the three survivors in inherited order as $(s_0,s_1,s_2)$.  The fixed
decoder unitary $W$ acts on those relabeled survivors and an ancilla $a$
initialized in $\lvert0\rangle$, independently of $d$, and satisfies
\begin{equation}
  \operatorname{Tr}_{s_1s_2a}\!\left[
    W\left(\operatorname{Tr}_d[V_{\mathrm{enc}}\rho
      V_{\mathrm{enc}}^\dagger]\otimes
      \lvert0\rangle\!\langle0\rvert_a\right)W^\dagger
  \right]=\rho
  \quad\text{on register } s_0\text{, for every input state }\rho.
  \label{eq:deletion-recovery-identity}
\end{equation}
The left side is the state of the remaining register $s_0$;
thus $s_0$ is the selected logical output.

For every tetrahedral input and each of the four positions, the circuit
prepares the input, applies $V_{\mathrm{enc}}$, parks the designated
carrier, and excludes that carrier from every later gate and routing path.
The designated $d$ is supplied only to the
offline compiler for parking, routing, and output binding; the logical decoder gate
sequence is the same for every $d$, while each submitted program, and so each hardware run, is compiled for its designated position. The decoder's logical ancilla is
initialized at circuit start and left untouched at the logical layer until
coherent, measurement-free decoding.
The decoded output is inverse-prepared and measured, while all other outputs,
including the parked carrier, are marginalized.  Parking and marginalization
therefore realize the subsystem partial trace in
\cref{eq:synthetic-deletion} within the ideal circuit model. The physical
carrier remains on the device.

Before hardware submission, the logical construction and its deterministic
gate lowering were checked in noiseless simulation for all 16 input-state by
deletion-position cells.  In every cell the selected-output success
probability agreed with one to within the prespecified numerical tolerance of
$3\times10^{-12}$.  The exact logical gate sequence is displayed in
\cref{fig:qec-circuit}.

Equivalently, let
$\operatorname{CR}_y(c\!\to\!t;\alpha)$ denote a controlled
$R_y(\alpha)$ rotation on target $t$ with control $c$, and let
$\theta_U\coloneqq 2\arccos(1/\sqrt3)$.  Reading left to right, on logical data wires
$\ell_0,\ldots,\ell_3$, the encoder block implementing
$V_{\mathrm{enc}}$ applies
\begin{equation*}
\begin{split}
&\operatorname{CR}_y(\ell_0\!\to\!\ell_1;\theta_U),\quad
\operatorname{CR}_y(\ell_1\!\to\!\ell_0;-\pi/2),\quad H_{\ell_2},\\
&\operatorname{CNOT}_{\ell_2\to \ell_1},\quad
\operatorname{CNOT}_{\ell_0\to \ell_3},\quad
\operatorname{CNOT}_{\ell_1\to \ell_3},\quad
\operatorname{CNOT}_{\ell_2\to \ell_0}.
\end{split}
\end{equation*}
On inherited survivors $(s_0,s_1,s_2)$ and the circuit-initialized ancilla $a$, the decoder
block implementing $W$ applies
\begin{equation*}
\begin{split}
&\operatorname{CNOT}_{s_2\to a},\quad
\operatorname{CNOT}_{s_1\to a},\quad
\operatorname{CNOT}_{s_0\to a},\quad
\operatorname{CNOT}_{a\to s_0},\quad
\operatorname{CNOT}_{a\to s_1},\\
&\operatorname{CNOT}_{a\to s_2},\quad
\operatorname{CNOT}_{s_0\to s_2},\quad
\operatorname{CNOT}_{s_1\to s_2},\quad
\operatorname{CR}_y(s_1\!\to\!s_0;\pi/2),\quad
\operatorname{CR}_y(s_0\!\to\!s_1;-\theta_U).
\end{split}
\end{equation*}

\section{Experimental setting}
\label{sec:design}

We accessed two processors through Amazon Braket: the sparse-connectivity
\IQM{} superconducting processor and IonQ Forte Enterprise 1.
IonQ programs specified five virtual qubits with advertised all-to-all
connectivity \citep{awsbraketionq}; the provider retained control of native
compilation and physical assignment and reported neither for either provider
task.  The experiments therefore provide separate diagnostics for the two
architectures. A controlled architecture comparison would require common
controls and comparable compilation and placement information.  Before examining the outcomes of each provider task, we fixed
the wires, operation order, measured bits, input-state order, success bits,
shot counts, and analysis endpoints.  All analyses use the raw returned outcomes, with no mitigation or
postselection applied in our analysis, every returned shot retained, and the original readout and mapping
conventions.  The retained records do not show whether the IonQ provider applied error mitigation or other processing before returning its aggregate histograms.

Decision rules were recorded locally in analysis plans before the
corresponding outcomes were inspected, without public preregistration or
external timestamping. The plan for the IQM recovery discovery specified
point summaries only.

IQM programs were routed and lowered offline, deterministically, into native
\texttt{PRX} and \texttt{CZ} gates on physical qubits inside full-verbatim
OpenQASM~3 boxes, so the submitted programs fixed the physical mapping.  A
device-capability document was retained before every IQM submission. Its
calibration fields lacked per-calibration timestamps, leaving their age
unknown; later pre-submission checks confirmed that the already fixed paths remained eligible.
\Cref{app:execution-provenance} records the device names and regions, compilation and source records, and provider-task chronology in
\cref{tab:task-provenance}. The recorded provider task-lifecycle spans measure elapsed service time, including
overhead.

All shot-level bounds and intervals in this work are
conditional model calculations: the Hoeffding bounds treat bounded shot
outcomes as independent, while the binomial bootstrap and Wilson intervals
add a Bernoulli/binomial shot model.  Shot independence is a modeling assumption: every analysis uses
per-program outcome counts, the IonQ provider returned only aggregate histograms,
and the published dataset contains no per-shot records, so correlation between
shots within a provider task was not examined.

Across eight provider tasks, all 624 programs and 101,600 requested shots completed.
The published dataset \citep{essayag2026reproducibility} links each program's reported counts to its submitted source; the provider records from which these counts were derived are held by the authors.

\section{Communication on fixed IQM routes}
\label{sec:communication}

This experiment asks whether differences between communication tasks recur
when the same native programs are executed again on the same physical routes.

\subsection{Design and decision rules}
\label{sec:communication-design}

The communication experiment used four initial qubits and two directed six-edge routes
per initial qubit, listed in \cref{tab:routes}.  The routes were selected and fixed
before the protocol outcomes were available.  Inference is conditional on these fixed diagnostic regions.

\begin{figure}[htbp]
  \centering
  \includegraphics[width=\textwidth,
    alt={The 53-node IQM graph with eight fixed routes. Initial qubit colors and shapes identify routes and their five-node recovery prefixes; concentric rings retain shared membership.}]{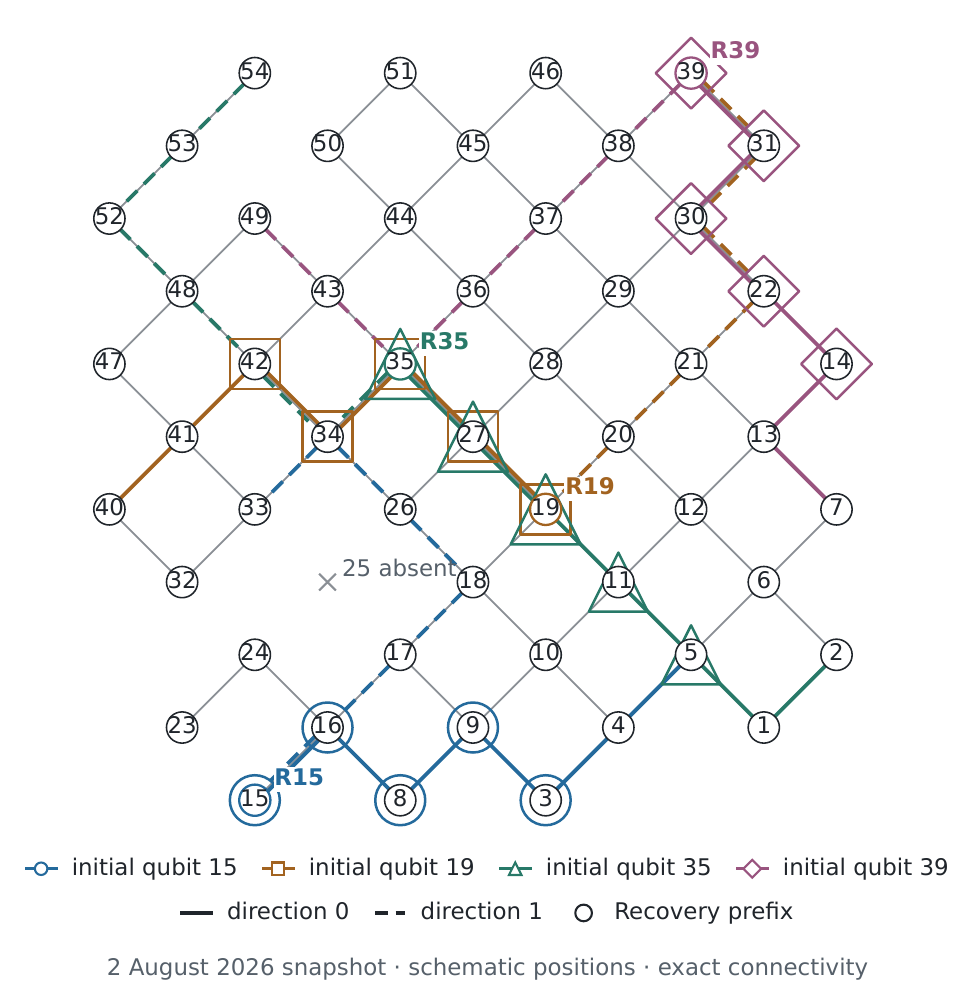}
  \caption{Eight fixed routes on the IQM capability graph captured on
  2 August 2026. Solid and dashed strokes distinguish the two ordered routes
  from each initial qubit; color, marker shape, and direct initial qubit labels identify their
  origin. Outer membership rings mark the five-node direction-zero prefixes
  used for recovery, with concentric rings at shared nodes. Routes that share an edge
  are drawn as parallel offset strokes. The ordered paths appear in
  \cref{tab:routes}. Connectivity is exact, but node positions are schematic.
  The service advertised 54 qubits; qubit~25 was absent from this captured
  graph, which also supplies the off-route edges.}
  \label{fig:frozen-routes}
\end{figure}

The initial qubits were spatial landmarks in IQM's published Crystal-54 geometry \citep{aws2025iqmemerald}:
15 and 39 are peripheral, while 19 and 35 are central. Routes were initially
selected by topology. When qubit 25 became unavailable, two routes were
repaired before any communication outcome: topology alone determined the direction-1 replacement
for initial qubit 15, while provider calibration fields resolved only a
remaining topology tie in choosing the direction-0 replacement for initial qubit 19. The pairs from initial qubits 35 and 39 were unchanged.
\Cref{app:route-selection} gives the complete deterministic selection and
repair rule. Let $G_{\mathrm{live}}$ denote the 53-node capability graph
captured on 2 August for the repaired communication routes.
\Cref{fig:frozen-routes}
places the eight final routes on $G_{\mathrm{live}}$;
\cref{fig:qec-placement-bindings} separately
shows the exact five-qubit recovery maps and carrier bindings.

Within each comparison, we pair the cells of the two protocols by directed route and corresponding
input-state or measurement-basis label.

\begin{table}[H]
  \centering
  \caption{The eight physical routes used in both communication executions.  Route
  length is counted in edges.}
  \label{tab:routes}
  \begin{tabular}{ccl}
    \toprule
    Initial qubit & Direction & Ordered physical route \\
    \midrule
    15 & 0 & $15$--$16$--$8$--$9$--$3$--$4$--$5$ \\
    15 & 1 & $15$--$16$--$17$--$18$--$26$--$34$--$33$ \\
    19 & 0 & $19$--$27$--$35$--$34$--$42$--$41$--$40$ \\
    19 & 1 & $19$--$20$--$21$--$22$--$30$--$31$--$39$ \\
    35 & 0 & $35$--$27$--$19$--$11$--$5$--$1$--$2$ \\
    35 & 1 & $35$--$34$--$42$--$48$--$52$--$53$--$54$ \\
    39 & 0 & $39$--$31$--$30$--$22$--$14$--$13$--$7$ \\
    39 & 1 & $39$--$38$--$37$--$36$--$35$--$43$--$49$ \\
    \bottomrule
  \end{tabular}
\end{table}

Each execution contained the 100 programs summarized in
\cref{tab:protocol-inventory}.  Every program received 200 shots, for 20,000
shots per execution.  Entanglement swapping was restricted to direction zero;
the other three families used both directions.  The discovery provider task was
created on 2 August 2026.  The exact temporal replication ran on 4 August 2026
with byte-identical submitted OpenQASM program sources, the same placements,
route and measurement order, and identical shots per program; each execution had
its own captured capability document, bootstrap seed, provider task identifier,
timestamps, and result records.  We analyze the two executions separately.

\begin{table}[H]
  \centering
  \caption{Program inventory in each of the two communication executions.}
  \label{tab:protocol-inventory}
  \begin{tabular}{lccc}
    \toprule
    Family & Physical routes & Cells per route & Programs \\
    \midrule
    Round-trip state echo & 8 & 4 input states & 32 \\
    Measurement-free teleportation & 8 & 4 input states & 32 \\
    Bell-state transfer & 8 & 3 Pauli bases & 24 \\
    Entanglement swapping & 4 & 3 Pauli bases & 12 \\
    \midrule
    Total & & & 100 \\
    \bottomrule
  \end{tabular}
\end{table}

The compilation resources in \cref{tab:protocol-resources} were recounted
directly from all 100 submitted native sources and the corresponding routed sources.  They contain 7,836
\texttt{PRX} and 3,800 \texttt{CZ} operations in total; every per-program
count agrees with the count recorded in the authors' program index.  Native \texttt{CZ}
depth is the as-soon-as-possible layer count of the submitted \texttt{CZ}
stream, preserving
the per-qubit \texttt{CZ} order while excluding single-qubit gates.

\begin{table}[H]
  \centering
  \caption{Per-program compilation resources for the communication families.
  Slash-separated \texttt{PRX} counts are ordered as $X/Y/Z$; all four
  tetrahedral-state settings have the same count.  Algorithmic \texttt{SWAP}
  is counted before native lowering.  Routing \texttt{SWAP} denotes an
  additional compiler-inserted macro beyond the algorithmic circuit.}
  \label{tab:protocol-resources}
  \begingroup
  \small
  \setlength{\tabcolsep}{4pt}
  \begin{tabular}{lrrrrrr}
    \toprule
    Family & Programs & \shortstack{Native\\\texttt{PRX}} & \shortstack{Native\\\texttt{CZ}}
      & \shortstack{\texttt{CZ}\\depth} & \shortstack{Algorithmic\\\texttt{SWAP}} & \shortstack{Routing\\\texttt{SWAP}} \\
    \midrule
    Round-trip state echo & 32 & 74 & 36 & 36 & 12 & 0 \\
    Bell-state transfer & 24 & $65/65/63$ & 31 & 19 & 10 & 0 \\
    Measurement-free teleportation & 32 & 112 & 55 & 45 & 17 & 0 \\
    Entanglement swapping & 12 & $29/29/27$ & 12 & 11 & 3 & 0 \\
    \bottomrule
  \end{tabular}
  \endgroup
\end{table}

The 28 one-sided route-performance endpoints were exactly eight RTSE and eight
measurement-free teleportation return scores compared with $2/3$, together with eight
Bell-state transfer and four direction-zero swapping overlaps compared with $1/2$;
an endpoint passes when its decision lower bound exceeds its reference.
Their multiplicity-controlled pass counts describe individual executions.
The four confirmatory targets below instead test temporal replication.
Using \cref{eq:teleportation-penalty}, define
$\widehat\Delta_{\mathrm{tel},r,j}$ for each of the eight routes,
$\overline\Delta_{\mathrm{tel},r}\coloneqq\frac12\sum_{j=0}^{1}
\widehat\Delta_{\mathrm{tel},r,j}$ for each initial qubit, and
$\overline\Delta_{\mathrm{tel}}\coloneqq\frac14\sum_{r\in\mathcal R}
\overline\Delta_{\mathrm{tel},r}$.  Also define
$\widehat\Delta_{\mathrm{swap},r}\coloneqq
\widehat F_{\mathrm{swap},r,0}-\widehat F_{\mathrm{Bell},r,0}$.
The 18-member two-sided contrast family was exactly the eight route-level
$\widehat\Delta_{\mathrm{tel},r,j}$ values, four $\overline\Delta_{\mathrm{tel},r}$ values, one for each initial qubit, the one aggregate
$\overline\Delta_{\mathrm{tel}}$, four route-level
$\widehat\Delta_{\mathrm{swap},r}$ values, and the aggregate
$\overline\Delta_{\mathrm{swap}}$ in \cref{eq:swapping-bell-contrast}.

Conditional on the 100 fixed program cells, we used nominal simultaneous
95\% binomial-bootstrap bounds: one-sided bounds for the 28 route-performance endpoints
and two-sided intervals for the 18 contrasts. The 95\% level applies separately to the endpoint family and to the contrast family. Protocol means within each execution
have pointwise intervals. Bell and swapping decisions additionally
use the smaller of the simultaneous bootstrap lower bound and a
Bonferroni--Wilson bound over the three Pauli components. These intervals
describe finite-shot uncertainty under the shot model of \cref{sec:design}.
\Cref{app:communication-intervals} gives the smoothing, resampling,
standardization and Wilson constructions.

After observing the discovery result, four confirmatory targets were
specified before the replication execution was submitted.  The primary
target was the route-39 direction-1 RTSE--teleportation contrast; it
replicated only if the lower limit of its simultaneous familywise interval
was greater than zero.  The secondary target was the aggregate
RTSE--teleportation contrast, with directions weighted equally within
each initial qubit and the four initial qubits then weighted equally; it used the same
rule.  The third target asked whether
route~39 direction~1 again combined an RTSE pass with a teleportation failure, the case in which RTSE alone would give false reassurance: it required an RTSE lower
bound above $2/3$, a teleportation lower bound not above $2/3$, and replication
  of the primary contrast.  Finally, the direction-zero
swapping--Bell contrast on the same routes in \cref{eq:swapping-bell-contrast} replicated only if
its simultaneous interval had a lower limit greater than zero.

\FloatBarrier
\subsection{Similar aggregate echo scores and changed route-level estimates}
\label{sec:results}

In discovery, 27 of the 28 prespecified route-performance endpoints passed:
all eight RTSE routes, seven of eight measurement-free teleportation routes, all eight
Bell-state transfer routes, and all four swapping routes.  All 28 passed in the
replication execution.  None of the four confirmatory replication targets in
\cref{sec:communication-design} met its rule.

\begin{table}[htbp]
  \centering
  \caption{Estimates from two executions of byte-identical
  submitted native program sources.  Parentheses give 95\% intervals.
  Intervals for individual
  family means are pointwise; intervals for the two displayed prespecified
  contrasts are simultaneous within their paired-contrast family.  The Bell
  row averages all eight directed routes. The final row averages paired
  swapping-minus-Bell differences on the four direction-zero routes.}
  \label{tab:snapshot-results}
  \small
  \begin{tabular}{@{}lcc@{}}
    \toprule
    Quantity & Discovery, 2 August & Replication, 4 August \\
    \midrule
    Round-trip state echo
      & $0.83844\;(0.82950,0.84738)$
      & $0.83969\;(0.83075,0.84862)$ \\
    Measurement-free teleportation
      & $0.79219\;(0.78239,0.80198)$
      & $0.82547\;(0.81619,0.83474)$ \\
    RTSE--teleportation
      & $0.04625\;(0.02624,0.06626)$
      & $0.01422\;(-0.00521,0.03364)$ \\
    Bell-state transfer
      & $0.77219\;(0.75709,0.78729)$
      & $0.79219\;(0.77762,0.80676)$ \\
    Entanglement swapping
      & $0.84375\;(0.82545,0.86205)$
      & $0.82938\;(0.81024,0.84851)$ \\
    Swapping--Bell, direction 0, same routes
      & $0.04313\;(0.00183,0.08442)$
      & $0.04063\;(-0.00188,0.08313)$ \\
    \bottomrule
  \end{tabular}
\end{table}

The RTSE means differed by $0.00125$ between executions (\cref{tab:snapshot-results}).
Across the eight fixed
routes, the route-level RTSE point estimates changed by between $-0.05250$ and
$+0.06125$; these descriptive changes have opposite signs and partly cancel in
the aggregate. Equivalence
would require a prespecified margin and a corresponding test.  The teleportation average
increased by $0.03328$, so the aggregate RTSE--teleportation contrast contracted from
$0.04625$ to $0.01422$ and its replication interval included zero.  The Bell-state
transfer mean over all eight routes increased by $0.02000$, but on the four
direction-zero routes Bell-state transfer and swapping both decreased, by
$0.01188$ and $0.01438$.  Their direction-zero contrast therefore changed only from
$0.04313$ to $0.04063$, but its replication interval included zero, so it did
not meet its prespecified replication rule.

The largest route-level discovery contrast was on route 39, direction 1, whose physical path
was $39$--$38$--$37$--$36$--$35$--$43$--$49$.  Its RTSE estimate was
$0.85875$, with simultaneous lower bound $0.82414$, whereas measurement-free
teleportation was $0.64750$, with lower bound $0.59991$.  The paired contrast
was therefore $0.21125$, with familywise interval
$[0.14989,0.27261]$.  In the exact replication the same route's contrast was
$0.04500$, with interval $[-0.01227,0.10227]$.  This replication interval crosses zero and remains compatible with a smaller positive
contrast.  The third target also failed: it required this
replication and a teleportation lower bound not above $2/3$, whereas
teleportation on this route passed its $2/3$ test in replication.
These replication rules are stringent: the primary target was the largest of
the eight route-level discovery contrasts, so selection alone favors some
contraction on repetition, and every target was judged with intervals
simultaneous over all 18 prespecified contrasts.  A failed rule therefore shows
that an effect did not recur at this stringency, not that it vanished.

\begin{figure}[!htbp]
  \centering
  \includegraphics[width=\textwidth,
    alt={Four route scatter plots show task scores that vary between executions despite close aggregate echo means. The caption explains the route symbols and reference lines.}]{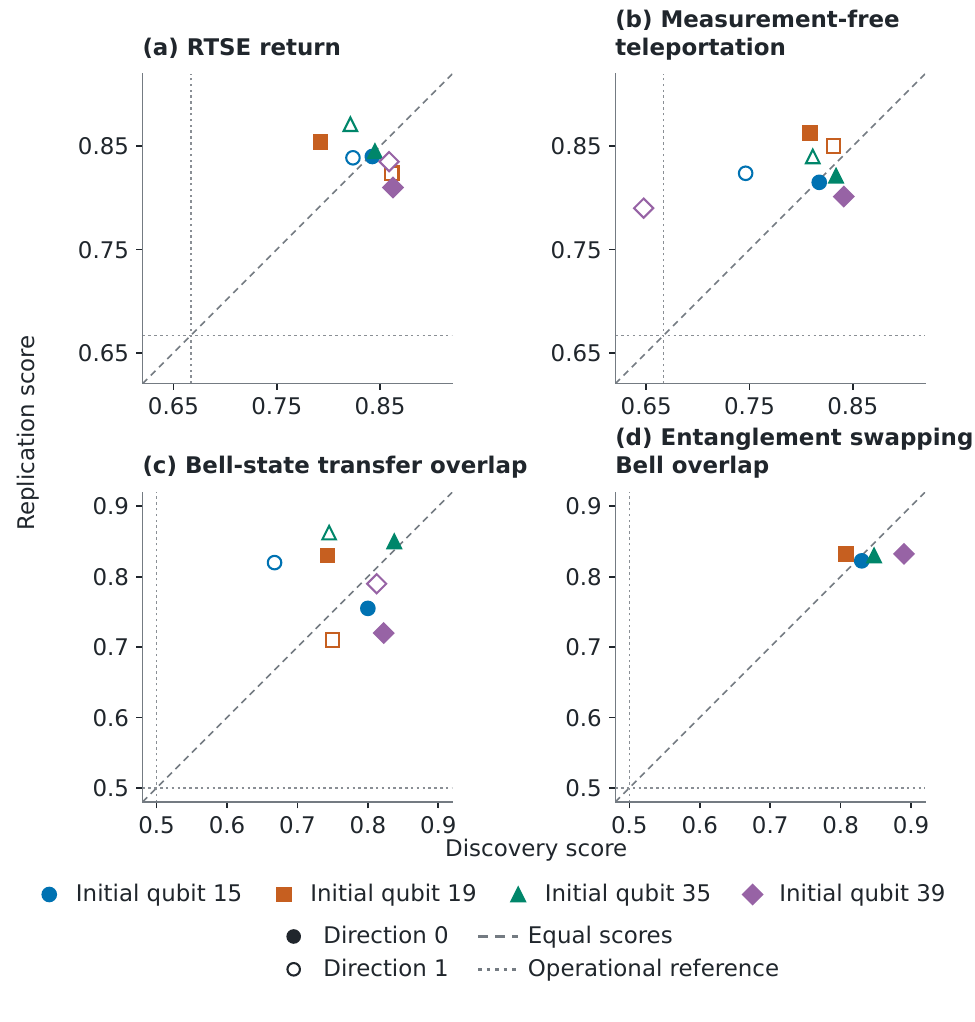}
  \caption{Route scores at discovery and temporal replication of byte-identical
  submitted native programs.  The route-39, direction-1 RTSE--teleportation
  contrast contracts, although aggregate echo means differ by only $0.00125$.
  Horizontal and vertical axes give discovery and replication scores;
  dashed diagonals mark equality.  Dotted operational references are $2/3$ for
  RTSE and teleportation (a,b), and $1/2$ for Bell overlap (c,d).
  Color and shape identify the physical initial qubit; filled and open markers denote
  directions 0 and 1.  Swapping (d) has direction 0 only.  All 28 markers are
  point estimates; the text gives the simultaneous intervals for route~39,
  direction~1.}
  \label{fig:route-replication}
\end{figure}
\Cref{fig:route-replication} places the selected contrast alongside the
route-resolved scores of all four families from both executions.

\section{Route-length response on IQM}
\label{sec:length-experiment}

The length experiment asks whether the two round-trip scores at the initial
qubit respond to transport distance, as a route diagnostic should, and whether
RTSE's score falls less than its predecessor's as a fixed route
is extended.

\subsection{Design and decision rules}
\label{sec:rtse-length-design}

The second question was selected prospectively as the directional hypothesis,
because we expected the simpler RTSE implementation to retain more
return success as the route grew. The idealized calculation in \cref{sec:rtse-length-results} shows that the opposite can also occur: return-leg dephasing lowers the ideal RTSE score while preserving the ideal predecessor marginal. RTSE keeps inverse preparation
at the initial qubit and reads one qubit, while carrying the arbitrary input on both
legs; the predecessor returns $\lvert0\rangle$ and reads every route qubit.
These circuit differences motivate an empirical comparison of the complete
workloads.

The length study used two vertex-disjoint geodesic route families,
\begin{align*}
  \mathcal P_{15}&\coloneqq(15,16,17,18,19,20,21,29,37,38,46),\\
  \mathcal P_{39}&\coloneqq(39,31,30,22,14,13,7,6,5,4,3).
\end{align*}
For $r\in\{15,39\}$, write
$\mathcal P_r^{(L)}$ for the prefix formed by the first $L+1$ vertices of $\mathcal P_r$.  Every prefix at
$L\in\{2,4,6,8,10\}$ has graph distance exactly
$L$ from its initial qubit in $G_{\mathrm{live}}$, the capability graph captured at
07:58~UTC on 2 August 2026 (\cref{sec:communication-design}).  Selection used topology, geodesic reach, vertex disjointness, and the
landmark initial qubits fixed in \cref{sec:communication-design}. Earlier IQM outcomes were available; route selection used only these structural criteria.
The routes, circuits, submitted orders, estimands, and analysis were fixed
before any length-study outcome was available; this plan is included in
the published dataset \citep{essayag2026reproducibility}.

The study used two windows, A and B, each a single provider task containing the same 96 programs at 200 shots per program.  Eight shared
$L=0$ cells crossed the two initial qubits with the four tetrahedral states.  At positive
length, 40 RTSE cells and 40 predecessor cells crossed two routes, five lengths,
and four states.  Eight additional one-way circuits tested the two
length-10 endpoints with the same four states.  These circuits prepared at the
initial qubit, transported once, applied the inverse preparation at the endpoint, and
measured there.
The predecessor's primary outcome was its marginal at the initial qubit; its
secondary joint outcome required the initial qubit and all returned prefix
ancillas to be zero.

The native inventory in each window contained 1,040 algorithmic \texttt{SWAP}
macros before lowering, 3,120 \texttt{CZ}, 6,432 \texttt{PRX}, and 336 terminal
measurements, with no compiler-inserted routing \texttt{SWAP}.  Window A used
a fixed order of eight blocks, one per route and input state, in which the
length-zero and one-way cells bracket five adjacent RTSE--predecessor pairs; between
blocks, the bracketing order and the first protocol of each pair alternate, and the order of
lengths rotates.  Window B used the exact reverse of this order; the same 96
native source strings, inputs, measurements, and 200-shot counts were retained.
The two windows ran about three hours apart on 14 August 2026.
Execution of B required technical completeness of A, irrespective of its
estimated effects.  Source identity was verified in both windows (\cref{app:length-source-bindings}).

Let $R_{P,w}(L)$ be the equal-weight mean of the probabilities of outcome zero at the initial qubit over the two routes and four states for protocol
$P\in\{\mathrm{RTSE},\mathrm{pred}\}$ in window $w\in\{A,B\}$,
where ``pred'' denotes the predecessor marginal at the initial qubit. The co-primary
length-drop and interaction estimands were
\begin{align}
  D_{P,w}&\coloneqq R_{P,w}(2)-R_{P,w}(10),\label{eq:length-drop}\\
  I_w&\coloneqq D_{\mathrm{pred},w}-D_{\mathrm{RTSE},w}.
  \label{eq:length-interaction}
\end{align}
A positive $I_w$ therefore means that RTSE's score at the initial qubit drops less than the predecessor's from
$L=2$ to $L=10$.  Equal-window estimands give the two windows equal weight. We estimate
these means and contrasts by replacing each success probability with
its raw observed cell proportion.

The confirmatory max-statistic family contained each of the three co-primary
estimands in window A, window B, and the equal-window analysis (nine members),
together with the one-way test score for each of two routes in each of two windows (four
members), for 13 members in total.  Each claim required its nominal, model-based
simultaneous one-sided 95\% lower bound to exceed the prespecified threshold
$0.05$, which we chose in advance as a study-specific
minimum effect of scientific interest.  A replicated co-primary claim additionally required the
corresponding decision in both windows and in the equal-window analysis, with
both one-way tests passing in each window.

For each one-way test, the estimand is its
tetrahedral success probability minus the input-independent average $1/2$:
if the endpoint output is independent of $s$, the zero first moment in
\cref{eq:tetrahedron-two-design} makes its mean
overlap with the four targets exactly $1/2$ under ideal preparation and readout.  Thus each one-way test requires, as its primary criterion, a
lower bound on success above $0.55$.
The same bound was secondarily compared with $2/3$, the operational
measure-and-prepare reference under the fixed-channel and trusted-SPAM
assumptions stated in \cref{sec:profile}.

The simultaneous bounds use a parametric max-statistic bootstrap that
retains the shared cells and the dependence between the predecessor's outcome zero at the initial qubit
and its all-zero outcome. They give nominal, model-based coverage conditional on these two windows
and fixed cells under the shot model in \cref{sec:design}; drift and
device-population inference require additional modeling.
The three $L=0$ loss contrasts are the equal-window drops from the shared $L=0$ mean to the $L=10$ scores of RTSE, the predecessor marginal and the predecessor all-zero event. These contrasts, the predecessor all-zero score,
and shape descriptions beyond the length drop are secondary or exploratory.
The length drop is the confirmatory shape question. The eight equal-window drops between adjacent
lengths, for RTSE and for the predecessor marginal, form a separate exploratory family with its own simultaneous one-sided 95\% lower bounds.
The complete joint-resampling construction appears in
\cref{app:length-intervals}.

\FloatBarrier
\subsection{Replicated length loss in both protocols}
\label{sec:rtse-length-results}

In each length window, all 96 programs completed and all 19,200 requested
shots were returned.  The one-way tests at both endpoints passed in both windows.
In the order (initial qubit, window) $(15,A),\allowbreak(39,A),\allowbreak(15,B),\allowbreak(39,B)$,
their four point estimates were $0.87125$, $0.89500$,
$0.87750$, and $0.85750$; the corresponding simultaneous lower bounds were
$0.84065$, $0.86672$, $0.84720$, and $0.82584$.  Thus each lower bound exceeded
the primary endpoint-sensitivity threshold $0.55$.  They also exceeded the
secondary two-thirds reference, under the fixed-channel and SPAM assumptions
stated in \cref{sec:profile}.

The co-primary length-drop results are given in
\cref{tab:rtse-length-endpoints}.  RTSE lost $0.17063$ from $L=2$ to $L=10$ in
window A and $0.20188$ in window B.  Both simultaneous lower bounds exceeded
the prespecified threshold $0.05$, as did the equal-window drop
$0.18625$.  The predecessor marginal also showed a replicated length
drop: $0.08063$ in A, $0.13125$ in B, and $0.10594$ under equal-window
weighting.
Descriptively, the RTSE length drop had the same positive sign in every
route--window stratum: $0.17875$ and $0.16250$ on $\mathcal P_{15}$ and
$\mathcal P_{39}$ in window A, and $0.18875$ and $0.21500$ in window B.

\begin{table}[H]
  \centering
  \small
  \caption{Prespecified $L=2$-to-$L=10$ endpoints.  Each entry is the estimate
  followed in parentheses by its nominal, model-based simultaneous one-sided
  95\% lower bound.  Passing requires the lower bound to exceed $0.05$.
  Positive interaction means that RTSE's score at the initial qubit drops less than the
  predecessor's; the observed interaction has the opposite sign.}
  \label{tab:rtse-length-endpoints}
  \begin{tabular}{>{\raggedright\arraybackslash}p{0.30\textwidth}ccc}
    \toprule
    Estimand & Window A & Window B & Equal-window \\
    \midrule
    RTSE drop
      & \shortstack{$0.17063$\\$(0.14105)$} & \shortstack{$0.20188$\\$(0.17032)$}
      & \shortstack{$0.18625$\\$(0.16457)$} \\
    \addlinespace[3pt]
    Predecessor marginal drop
      & \shortstack{$0.08063$\\$(0.05649)$} & \shortstack{$0.13125$\\$(0.10427)$}
      & \shortstack{$0.10594$\\$(0.08787)$} \\
    \addlinespace[3pt]
    Protocol-by-length interaction
      & \shortstack{$-0.09000$\\$(-0.12822)$} & \shortstack{$-0.07063$\\$(-0.11223)$}
      & \shortstack{$-0.08031$\\$(-0.10859)$} \\
    \bottomrule
  \end{tabular}
\end{table}

The interaction in \cref{eq:length-interaction} was negative in both windows
and in the equal-window analysis, so the prespecified RTSE retention claim
was not supported.  Because the confirmatory test concerned only the positive directional alternative,
the observed negative interaction is descriptive evidence about the opposite
direction. It compares complete workloads with different return states and
terminal measurements; separating their physical contributions requires
additional controls.

An idealized calculation for the return channel isolates one difference in sensitivity.
Assume perfect outward transport, ideal preparation, inverse preparation and
readout, and the same fixed, input-independent qubit channel on the return leg
of both protocols.  Write its Bloch action as
$\mathbf v\mapsto M\mathbf v+\mathbf t$.  For RTSE, the return overlap is
$p_s(0)=(1+\mathbf v_s^{\mathsf T}M\mathbf v_s+
\mathbf v_s^{\mathsf T}\mathbf t)/2$.
The tetrahedral moments in \cref{eq:tetrahedron-two-design} therefore give the ideal RTSE score and the ideal predecessor marginal,
\begin{equation}
  R_{\mathrm{RTSE}}^{\mathrm{ideal}}
    =\frac12+\frac{\operatorname{Tr}M}{6},
  \qquad
  R_{\mathrm{pred}}^{\mathrm{ideal}}
    =\frac{1+M_{zz}+t_z}{2}.
  \label{eq:return-channel-comparison}
\end{equation}
The second expression follows because remote inverse preparation makes the
predecessor's returning state $\lvert0\rangle$.  The predecessor's joint all-zero score also depends on the route ancillas.
For example, a dephasing return channel with
$M=\operatorname{diag}(\eta,\eta,1)$, $\mathbf t=0$ and
$0\leq\eta\leq1$ gives $R_{\mathrm{RTSE}}^{\mathrm{ideal}}=(2+\eta)/3$
and $R_{\mathrm{pred}}^{\mathrm{ideal}}=1$.
This example shows how loss of coherence on the return leg can affect the two
marginals differently.  It does not identify the noise in the IQM data:
noisy outward transport, input-dependent compilation and SPAM may also affect the
observed comparison, and noisy outward transport need not leave the
predecessor's remotely inverse-prepared state equal to $\lvert0\rangle$.

\begin{figure}[H]
  \centering
  \includegraphics[width=\textwidth,
    alt={Two curve panels and a length-drop plot show larger echo loss and negative interaction in both windows. The caption explains the one-sided bounds and decisions.}]{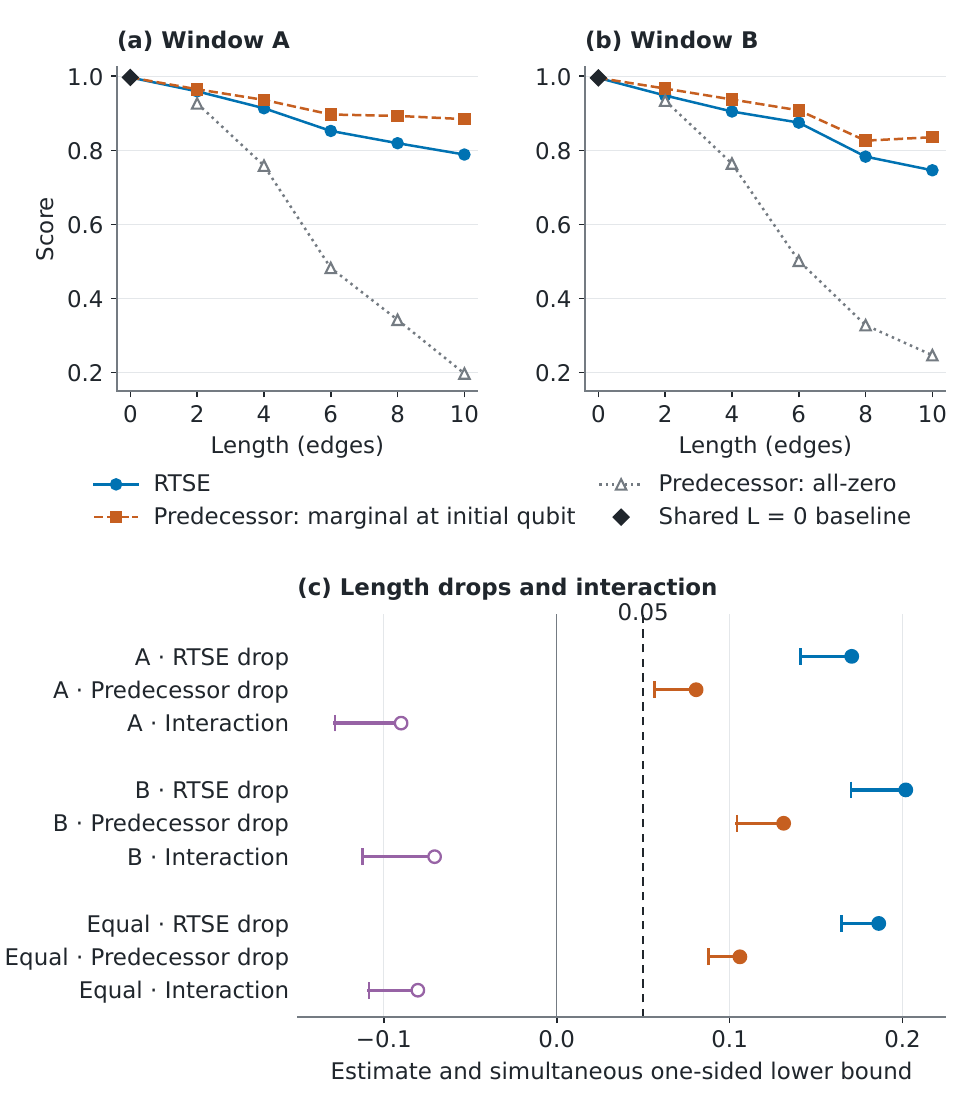}
  \caption{Length-dependent loss in two windows.  Panels (a,b) distinguish
  the scores at the initial qubit from the predecessor's all-zero score.  In (c), circles mark
  the nine $L=2$-to-$L=10$ estimates and left caps their nominal, model-based
  simultaneous one-sided 95\% lower bounds.
  The four one-way tests in the same family are omitted here.
  Filled circles meet the prespecified lower-bound criterion $>0.05$; open circles
  do not.  The dashed line marks $0.05$, and ``Equal'' denotes equal-window
  weighting.  Values appear in \cref{tab:rtse-length-endpoints}.}
  \label{fig:rtse-length-results}
\end{figure}

In \cref{fig:rtse-length-results}, each positive-length curve point is the
equal-weight mean over two fixed routes and four tetrahedral states.  RTSE
and predecessor marginals at the initial qubit are the co-primary protocol scores; the
predecessor all-zero event includes the wider terminal check and is secondary.
The shared $L=0$ local baseline is shown by a black diamond.  Connected
segments join the observed means at prescribed lengths. Panel (c)
reports the simultaneous uncertainty bounds for these two windows and routes.

The full curves in \cref{fig:rtse-length-results} show why the marginal at the initial qubit
and the predecessor's joint all-zero outcome must be distinguished.  Under equal-window
weighting, RTSE decreased from $0.95375$ at $L=2$ to $0.76750$ at $L=10$,
whereas the predecessor marginal decreased from $0.96594$ to $0.86000$.
The predecessor all-zero score fell much more sharply, from $0.93063$ to
$0.22219$, because every returned route ancilla enters that event.  The shared
$L=0$ mean was $0.99625$.

In the equal-window exploratory family, all four adjacent RTSE differences
and their within-family lower bounds were positive, whereas the predecessor
marginal's final $L=8$-to-$L=10$ difference was $-0.00031$, with lower bound
$-0.02186$.

\section{Deletion recovery on IQM}
\label{sec:iqm-recovery}

We next test repeatability of the recovery workload on four fixed physical
placements. The first execution supplies a discovery observation; the second
uses a decision rule fixed before its outcomes were available.

\subsection{Design and decision rules}

For the initial placement, index the data wires of \cref{fig:qec-circuit} as $\ell_0,\ldots,\ell_3$
and identify the decoder ancilla $a$ with logical wire $\ell_4$.
For each initial qubit $r\in\mathcal R$, define the initial recovery placement by
\begin{equation}
  \lambda_r(\ell_k)\coloneqq q_{r,0,k},\qquad k\in\{0,1,2,3,4\};
  \label{eq:qec-placement-map}
\end{equation}
thus the five logical wires inherit the first five physical vertices of the
direction-zero route, in route order.
Routing may move the decoded output before its fixed terminal measurement.

\begin{figure}[htbp]
  \centering
  \includegraphics[width=\textwidth,
    alt={Four five-qubit IQM placements with initial logical assignments. Each deletion row lists the parked physical carrier and final readout carrier; the values also appear in the body.}]{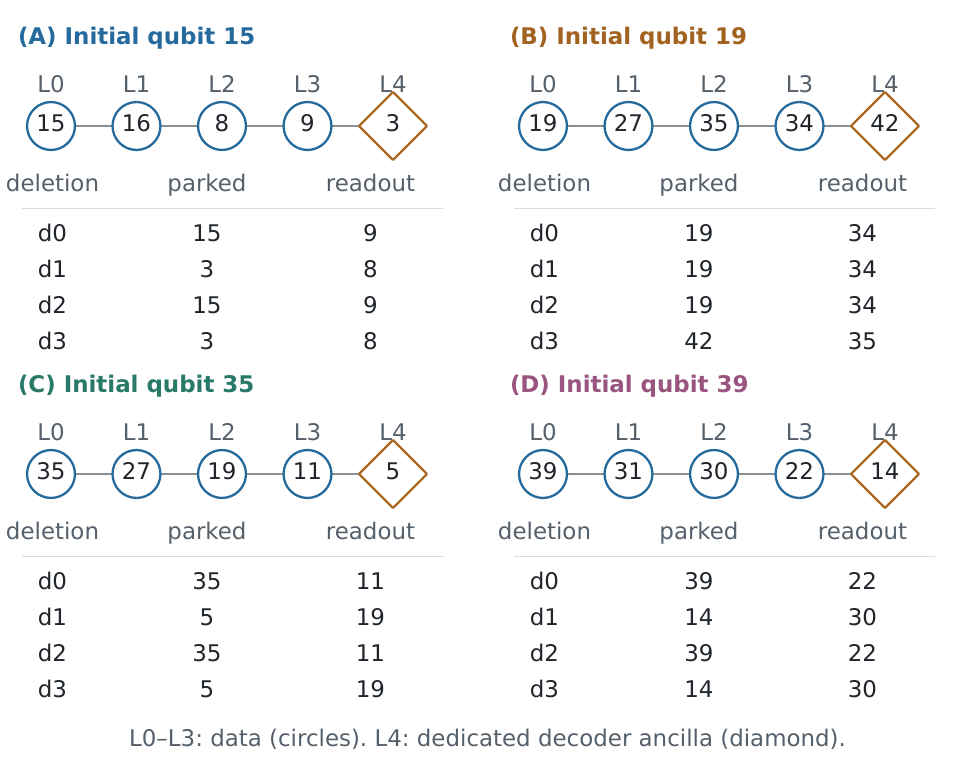}
  \caption{Initial placements and final carrier bindings for IQM recovery.
  Numbers are physical qubit identifiers; the labels \textsf{L0}--\textsf{L3} mark the initial data
  wires $\ell_0,\ldots,\ell_3$, and the diamond \textsf{L4} marks the dedicated decoder ancilla $\ell_4$. For
  each designated deletion position $d=0,\ldots,3$ (rows \textsf{d0}--\textsf{d3}), the table lists the physical carrier
  parked and excluded from subsequent gates and the carrier measured after
  recovery. Routing can change which logical state a physical carrier holds,
  so these boundary and readout bindings differ from the initial map.}
  \label{fig:qec-placement-bindings}
\end{figure}

The discovery execution ran on 6 August 2026.  Four isomorphic five-qubit
paths were inherited directly from the prospectively fixed protocol regions.  Their initial logical-to-physical maps
are shown in \cref{fig:qec-placement-bindings,tab:qec-placements}.  Recovery retained these maps, with live checks limited to connectivity and
calibration-field coverage.  The analysis plan of the communication discovery also set an outcome-based condition for proceeding to this experiment: at least four of the eight RTSE route-performance endpoints had to pass, at two or more initial qubits, and at least two initial qubits each needed one route on which the RTSE, measurement-free teleportation, and Bell-state transfer endpoints all passed.  In discovery all eight RTSE endpoints passed and every initial qubit had such a route, so the condition excluded no placement.  We cannot establish whether the decision to submit the recovery experiment depended on the condition; this leaves open only whether recovery would have been run had it failed.  Each placement contributed 16
recovery cells, one for each combination of tetrahedral input and designated deletion
position, and four controls that apply the encoder followed by its inverse.  Each of the two 80-program
provider tasks used 100 shots per program, for 8,000 shots per execution.

The discovery submission grouped programs by placement.  Before the second
execution on 7 August 2026, we fixed a cryptographically seeded balanced
interleave of placements, states, deletion positions, and controls.  The schedule used the first seed drawn.  Individual native programs,
placements, measurement bindings, success bits, and shot counts were otherwise
held fixed.  This randomization determines the submitted list; the provider's
chronological execution order within the program-set provider task remains unknown.
The replication schedule was fixed after discovery outcomes were known and
before the pre-submission check and the replication results were available.

\begin{table}[H]
  \centering
  \caption{Fixed five-qubit placements for the recovery workload.  The listed
  order maps logical wires $\ell_0,\ldots,\ell_4$ to physical qubits; $\ell_4$ is
  the dedicated decoder ancilla, initialized at circuit start and untouched at
  the logical layer until decoding. Placements B and C share qubits 19, 27 and 35.}
  \label{tab:qec-placements}
  \begin{tabular}{cll}
    \toprule
    Initial qubit & Label & Logical-to-physical map \\
    \midrule
    15 & placement A & $(15,16,8,9,3)$ \\
    19 & placement B & $(19,27,35,34,42)$ \\
    35 & placement C & $(35,27,19,11,5)$ \\
    39 & placement D & $(39,31,30,22,14)$ \\
    \bottomrule
  \end{tabular}
\end{table}

In the table below, each entry is ``parked physical
carrier $\to$ measured physical carrier'' after the designated deletion:
\begin{center}
\begin{tabular}{cllll}
  \toprule
  Placement & $d=0$ & $d=1$ & $d=2$ & $d=3$ \\
  \midrule
  A (initial qubit 15) & $15\to9$ & $3\to8$ & $15\to9$ & $3\to8$ \\
  B (initial qubit 19) & $19\to34$ & $19\to34$ & $19\to34$ & $42\to35$ \\
  C (initial qubit 35) & $35\to11$ & $5\to19$ & $35\to11$ & $5\to19$ \\
  D (initial qubit 39) & $39\to22$ & $14\to30$ & $39\to22$ & $14\to30$ \\
  \bottomrule
\end{tabular}
\end{center}

The panel labels give circuit-start assignments.  Encoding, routing, and pre-discard parking determine which physical carrier
holds logical wire $\ell_d$ at the discard boundary.
For example, in placement A with $d=1$, qubit 3 starts with the
decoder ancilla $\ell_4$ but holds the designated data wire $\ell_1$ at the discard
boundary; qubit 8 is measured after recovery.  The lower grid of \cref{fig:qec-placement-bindings} and the table above specify the boundary and readout
bindings after those compiled moves.

Before the discard boundary, the deterministic compiler moved the designated wire $\ell_d$, when necessary, to an end qubit of the five-qubit path (parking), leaving the four remaining wires on a connected four-qubit path.  At $d=1$ it parked $\ell_1$ at initial qubit 19 for placement B but at the starting qubit of $\ell_4$ for A, C and D (3, 5 and 14), with identical operation counts in \cref{tab:qec-resources}.
After the discard boundary, the parked carrier was excluded from every
native gate and routing path and was marginalized at output, which realizes
the partial trace of \cref{eq:synthetic-deletion} in the ideal circuit model.

Routing and decoding determined the recovered output's physical readout
carrier, listed for every placement and deletion position in the table above; it is part of the fixed measurement binding.  Because both codewords are permutation invariant, the ideal task is the same for every
deletion position; comparisons by deletion position therefore reflect physical differences,
including the routing burden in \cref{tab:qec-resources} and the readout carrier.

The compilation burden is reported in \cref{tab:qec-resources}.  All four
placements have the same counts for a fixed circuit family.  A routing
\texttt{SWAP} records a compiler-inserted macro before lowering into the
device's native gates.

\begin{table}[H]
  \centering
  \caption{Per-program operation counts for the recovery workload after
  placement and native compilation.}
  \label{tab:qec-resources}
  \begin{tabular}{lrrr}
    \toprule
    Circuit family & \texttt{CZ} & \texttt{PRX} & Routing \texttt{SWAP} \\
    \midrule
    Encoder followed by inverse & 31 & 74 & 5 \\
    Deletion position 0 & 65 & 141 & 15 \\
    Deletion position 1 & 68 & 147 & 16 \\
    Deletion position 2 & 59 & 129 & 13 \\
    Deletion position 3 & 68 & 147 & 16 \\
    \bottomrule
  \end{tabular}
\end{table}

For states $s\in\{0,1,2,3\}$ and deletion
positions $d\in\{0,1,2,3\}$, let $\widehat p_{r,s,d}(0)$ be the observed fraction
of shots in the corresponding recovery cell whose selected measured output bit is zero
after inverse preparation.  Let $\widehat p^{\mathrm{ctrl}}_{r,s}(0)$ denote
the analogous fraction for the short control that applies the encoder followed by its inverse.  The
primary recovery estimator is
\begin{equation}
  \Rrec\coloneqq\frac{1}{64}\sum_{r\in\mathcal R}
                     \sum_{s=0}^{3}\sum_{d=0}^{3}
                     \widehat p_{r,s,d}(0).
  \label{eq:qec-primary}
\end{equation}
A separate control of this kind is run for every placement and input
state,
\begin{equation}
  \Rctrl\coloneqq\frac{1}{16}\sum_{r\in\mathcal R}\sum_{s=0}^{3}
                    \widehat p^{\mathrm{ctrl}}_{r,s}(0).
  \label{eq:qec-control}
\end{equation}
The control measures input return through the shorter encoding sequence on
the same placement.  Its contrast with recovery combines the depth difference with the other
differences between the two complete workloads.  The recovery score measures
the processor's execution of the complete workload: preparation, encoding,
the circuit implementation of discard, decoding, inverse preparation, and readout.

The discovery plan fixed these point estimands, the four placement means, the
four deletion-position means, and the minimum of those four deletion-position means. Confidence intervals and
hypothesis testing were outside that plan, so discovery uncertainty
intervals are post hoc. The short control's descriptive pooled Wilson
interval uses a homogeneous-cell shot model.

Each of the 64 recovery cells contains 100 shots, so $\Rrec$ also equals
the total recovery success count divided by 6,400. Before the second recovery
execution, a local analysis plan fixed the one-sided error level $\alpha$,
the pooled recovery shot count $n$, and the Hoeffding lower bound \citep{hoeffding1963probability}
\begin{equation}
  L_{\mathrm H}
  \coloneqq\Rrec-\sqrt{\frac{\log(1/\alpha)}{2n}},
  \qquad n\coloneqq 6400,\quad \alpha\coloneqq 0.025,
  \label{eq:qec-hoeffding}
\end{equation}
and declared confirmation only if $L_{\mathrm H}>2/3$.  At the fixed shot
count this required at least 4,376 recovery successes. The plan was recorded locally before result
retrieval, without public preregistration or external timestamping.
The bound treats the fixed-cell shots as independent bounded trials (\cref{sec:design}).

The prespecified secondary analysis pairs the 64 replication-minus-discovery
cell changes and reports their conditional shot-noise interval, together with
a sensitivity interval based on the 16 placement-by-state blocks. The latter
retains the four deletion positions within each block. These summaries and the recovery-minus-control gap are secondary,
conditional descriptions of the fixed placements. The primary decision uses
the recovery bound alone. The detailed procedures are given in \cref{app:recovery-intervals}.

A prespecified secondary contrast between peripheral and central placements
was excluded because a metadata error had mislabeled the two groups
(\cref{app:excluded-spatial-summary}); placement-resolved,
state-resolved, deletion-resolved and primary results are unaffected.

\FloatBarrier
\subsection{Deletion recovery is not confirmed on replication}

All 80 programs completed in both recovery executions.  In discovery, 4,726
of 6,400 recovery shots returned the prepared input:
\begin{equation}
  \Rrec^{(\mathrm{disc})}=0.7384375.
\end{equation}
This point estimate was $0.07177$ above the two-thirds reference.  The
exploratory post-hoc 95\% placement-by-state block interval was
$[0.71703,0.75859]$.  The 16 short
controls succeeded in 1,395 of 1,600 shots, giving
$\Rctrl^{(\mathrm{disc})}=0.871875$ and a recovery-minus-control gap of
$-0.1334375$.

In the replication, only 3,994 of 6,400 recovery shots
succeeded:
\begin{equation}
  \Rrec^{(\mathrm{rep})}=0.6240625,
  \qquad L_{\mathrm H}=0.607086.
\end{equation}
The descriptive pooled Wilson interval was $[0.61212,0.63585]$.  The
confirmatory rule required at least 4,376 successes, so the replication missed its
prespecified threshold by 382 successes and did not confirm the discovery point
estimate's excess over $2/3$.  The controls succeeded in 1,312 of 1,600 shots, giving
$\Rctrl^{(\mathrm{rep})}=0.820000$ and a descriptive Wilson interval of
$[0.80042,0.83805]$.

The matched recovery change was
\begin{equation}
  \Rrec^{(\mathrm{rep})}-\Rrec^{(\mathrm{disc})}=-0.114375.
  \label{eq:qec-temporal-change}
\end{equation}
Its prespecified 95\% conditional shot-noise interval was
$[-0.13031,-0.09859]$; the 95\% 16-block placement-by-state sensitivity interval was
$[-0.14219,-0.08844]$.  Of the 64 matched recovery cells, 58 had lower
observed return proportions in replication, one was equal, and five had higher
observed proportions.  The control change was $-0.051875$,
so the recovery-minus-control gap became $0.0625$ more negative, with a
conditional 95\% difference-of-gaps interval of $[-0.09156,-0.03328]$.

\begin{figure}[!t]
  \centering
  \includegraphics[width=\textwidth,
    alt={Horizontal recovery intervals place the replication mean and its confirmatory lower bound below two-thirds. The caption distinguishes the two uncertainty summaries.}]{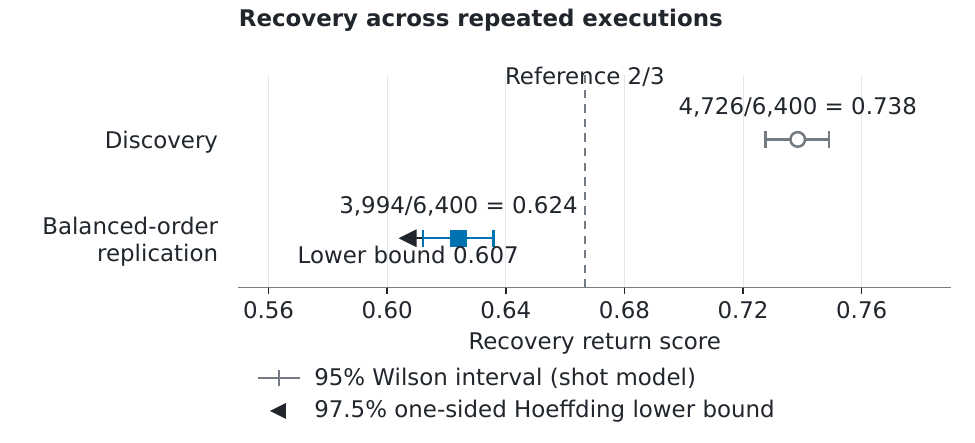}
  \caption{Deletion recovery is not confirmed on replication.  The
  horizontal axis gives the recovery return score; the two rows show the
  placement-blocked discovery and the balanced-order replication.  The open circle and filled square
  mark means from 64 fixed recovery cells with 100 shots each.  Capped bars
  are descriptive pooled 95\% Wilson shot-model intervals for the mean return probability.  The left-pointing
  triangle is the replication's prespecified 97.5\% one-sided Hoeffding lower bound on that probability;
  only this bound enters the confirmatory rule.  Both it and the replication
  mean lie below the dashed $2/3$ reference.  These shot-model summaries assume independent outcomes. The text reports
  the separate exploratory discovery placement-by-state block interval.
  ``Balanced order'' describes the submitted list; provider execution
  chronology remains unknown.}
  \label{fig:qec-confirmatory}
\end{figure}
\Cref{fig:qec-confirmatory} separates the descriptive two-sided intervals
from the one-sided bound that governed the prespecified replication decision.

The change between executions, which differed in date and submitted order, was distributed across every reported stratum
(\cref{tab:qec-results}).  Placement B showed the largest decline.  All four placement
means and all four deletion-position means were lower in replication.

\begin{table}[!htb]
  \centering
  \small
  \caption{Recovery return score by fixed placement and deletion
  position.  Change is replication minus discovery.  Each displayed stratum
  contains 1,600 recovery shots per execution.}
  \label{tab:qec-results}
  \begin{tabular}{lrrr@{\hspace{0.7em}}lrrr}
    \toprule
    Placement & Discovery & Replication & Change
      & Deletion & Discovery & Replication & Change \\
    \midrule
    Placement A & 0.76438 & 0.67250 & $-0.09188$
      & 0 & 0.73188 & 0.64250 & $-0.08938$ \\
    Placement B & 0.75813 & 0.56313 & $-0.19500$
      & 1 & 0.74813 & 0.59750 & $-0.15063$ \\
    Placement C & 0.71625 & 0.60813 & $-0.10813$
      & 2 & 0.76438 & 0.67375 & $-0.09063$ \\
    Placement D & 0.71500 & 0.65250 & $-0.06250$
      & 3 & 0.70938 & 0.58250 & $-0.12688$ \\
    \bottomrule
  \end{tabular}
\end{table}

\begin{figure}[p]
  \centering
  \includegraphics[width=\textwidth,
    alt={Four placement strips pair 64 recovery cells across executions: 58 decrease, one ties, and five increase. The caption explains the categorical ordering and exact proportions.}]{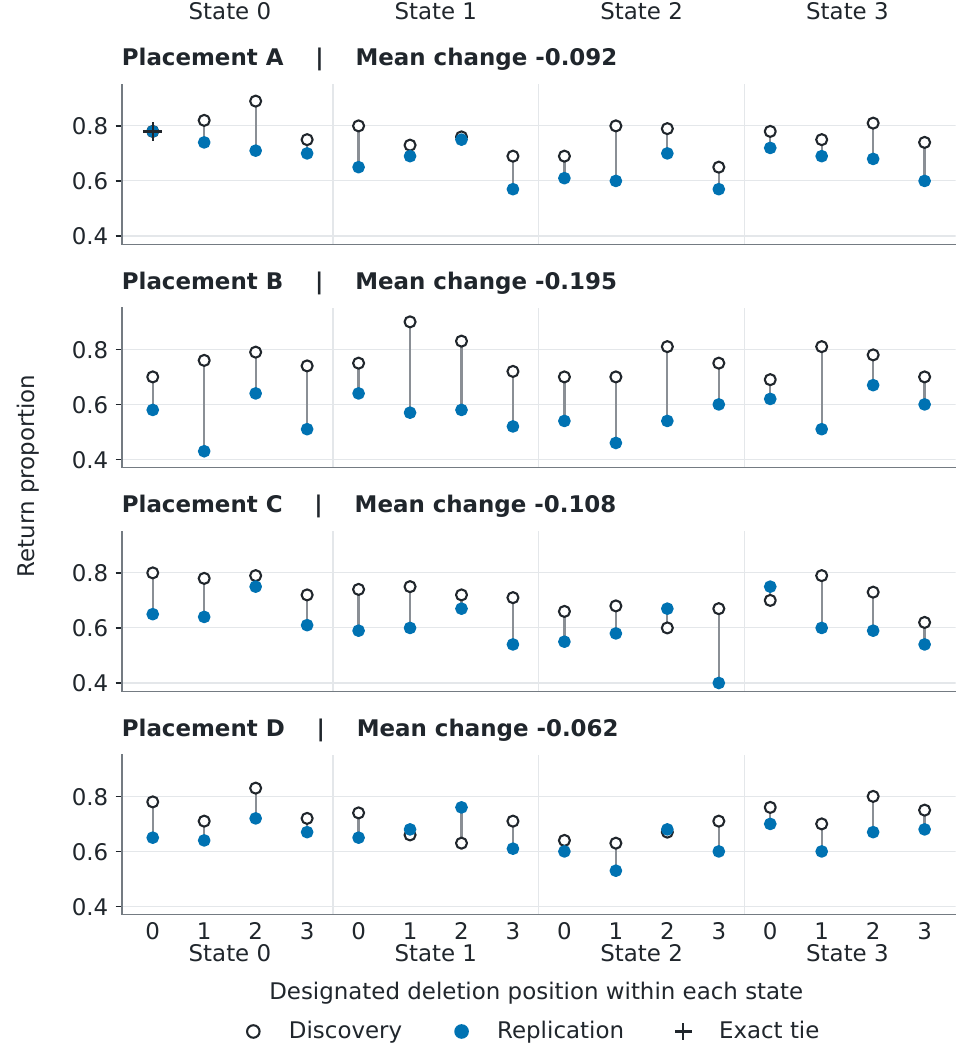}
  \caption{The recovery decline spans the fixed placements.  Four placement
  strips display all 64 matched cells: horizontal positions group input
  states $s=0,1,2,3$, then designated deletion positions $d=0,1,2,3$.
  Each gray segment joins
  an open discovery circle to a filled replication circle at the exact
  100-shot return proportions; a plus marks the sole exact tie.
    ``Mean change'' averages the 16 within-placement
  replication-minus-discovery differences.  Fifty-eight descending pairs
  and lower means at every placement show the later execution's broad decline.}
  \label{fig:qec-paired-cells}
\end{figure}

The replication list interleaved the placements, removing the discovery list's
placement blocking.

All four tetrahedral-state means also declined.  In state order, the two
vectors were
\[
\begin{aligned}
  \text{discovery:}&\quad(0.77250,0.74000,0.69688,0.74438),\\
  \text{replication:}&\quad(0.65125,0.62938,0.57688,0.63875).
\end{aligned}
\]
The full 64-cell pairing in
\cref{fig:qec-paired-cells} shows the decline across states and deletion positions.

\section{Deletion recovery on IonQ}
\label{sec:ionq-experiment}

The IonQ experiment asks whether the processor distinguishes correct recovery
from a control that replaces the decoder by its adjoint, and whether the same decisions
recur in a second window.

\subsection{Design and decision rules}
\label{sec:ionq-design}

The separate IonQ experiment reused the same four tetrahedral
inputs and all four designated deletion positions, giving 16 recovery cells
in each of two separately submitted windows.
Every recovery cell was paired with a deliberately incorrect
negative control, called the matched control below, having the same input state and deletion position.  In this
control, the correct coherent decoder $W$ was replaced by its adjoint while the
encoder, circuit implementation of discard, inverse input preparation, and
measured-output convention were retained.  Taking the adjoint reverses the
decoder operation order and the signs of its parameterized rotations;
the comparison tests the correct decoder against this particular
prespecified control, which combines both changes. Both circuits are encoded workloads, so this
comparison estimates a decoder-control contrast.

Noiseless statevector evaluation of the exact submitted QASM gives recovery
probability one in all 16 cells and a control mean of $0.473158$
(cell range $0.378553$--$0.567763$), hence an ideal mean separation of
$0.526842$. This is the ideal separation for the chosen circuit pair;
the hardware decisions use the prospectively fixed bounds below.

The ideal control map can also be calculated directly from the codewords and
the gate sequence in \cref{fig:qec-circuit}.  The three survivors retain their
inherited order $(s_0,s_1,s_2)$, the added ancilla $a$ starts in
$\lvert0\rangle$, and the selected output is $s_0$.  For a basis value $b\in\{0,1\}$ of the deleted carrier and an output basis label
$e\in\{0,1\}^3$ on $(s_1,s_2,a)$, the control has Kraus operators
\begin{equation*}
  K^{(d)}_{b,e}\coloneqq
  (I_2\otimes\langle e\rvert_{s_1s_2a})W^\dagger
  \bigl[(\langle b\rvert_d V_{\mathrm{enc}})
    \otimes\lvert0\rangle_a\bigr].
\end{equation*}
Here $\langle b\rvert_d$ contracts the deleted carrier.  The permutation
invariance of both codewords makes the channel
$\Phi_{\mathrm{adj}}(\rho)\coloneqq\sum_{b,e}K^{(d)}_{b,e}\rho
(K^{(d)}_{b,e})^\dagger$ the same for every $d$.  Substituting the printed
gates and codewords on the operator basis gives
\begin{equation*}
  \Phi_{\mathrm{adj}}(I_2)=I_2+\frac{1+2\sqrt6}{18}Z,
  \quad \Phi_{\mathrm{adj}}(X)=\Phi_{\mathrm{adj}}(Y)=0,
  \quad \Phi_{\mathrm{adj}}(Z)=\frac{1-\sqrt6}{9}Z.
\end{equation*}
Thus the decoded qubit, before input-dependent inverse preparation, has Bloch
action
\begin{equation}
  (x,y,z)\longmapsto
  \left(0,0,\frac{1+2\sqrt6}{18}+\frac{1-\sqrt6}{9}z\right).
  \label{eq:ionq-adjoint-map}
\end{equation}
The tetrahedral moments in \cref{eq:tetrahedron-two-design} then give, as in the first expression of \cref{eq:return-channel-comparison}, the exact ideal control mean
$R_{\mathrm{adj}}^{\mathrm{ideal}}=
\frac12+\frac{1-\sqrt6}{54}=\frac{28-\sqrt6}{54}$,
recovering the simulated value above.

Before lowering, the recovery and control circuits had logical depths
18 and 16, respectively: the as-soon-as-possible layer counts of their logical gate sequences, with each controlled $R_y$ one layer and $U_s$ and $U_s^\dagger$ two layers each.  After deterministic lowering of controlled
rotations, each circuit had the same submitted gate-type counts: 20
\texttt{CNOT}, 10 \texttt{RY}, two \texttt{RZ}, and one \texttt{H}
operation.  An as-soon-as-possible schedule of the submitted operation lists,
assigning one layer to every operation and respecting wire conflicts and list
order, gives greedy submitted depths 28 and 27.
The matching is at the level of submitted gate-type counts. Rotation
parameters and depths differ, and duration is unmatched. Provider-native
resources and physical placements remain unreported.
Four short controls that apply the encoder followed by its inverse, one for
each input state, contained 16 \texttt{CNOT}, 10 \texttt{RY}, two
\texttt{RZ}, and two \texttt{H} operations.
They test input return through the shorter sequence of
preparation, encoder, inverse encoder, and readout on the same declared
virtual qubits.
Concretely, virtual qubits 0--4 start in $\lvert0\rangle$, with qubit 4 retained
as idle padding.  The short circuit applies $U_s$ on qubit 0, the four-wire encoder
circuit in \cref{fig:qec-circuit}, its exact reversed inverse, and
$U_s^\dagger$ on qubit 0, in that order.  It measures only virtual qubit 0 and scores
the terminal bit value 0 as success.

\begin{figure}[!t]
  \centering
  \includegraphics[width=\textwidth,
    alt={IonQ role diagram, two-window circuit counts, and recovery/control sequences on five virtual qubits. The diagram identifies logical roles on submitted virtual qubits.}]{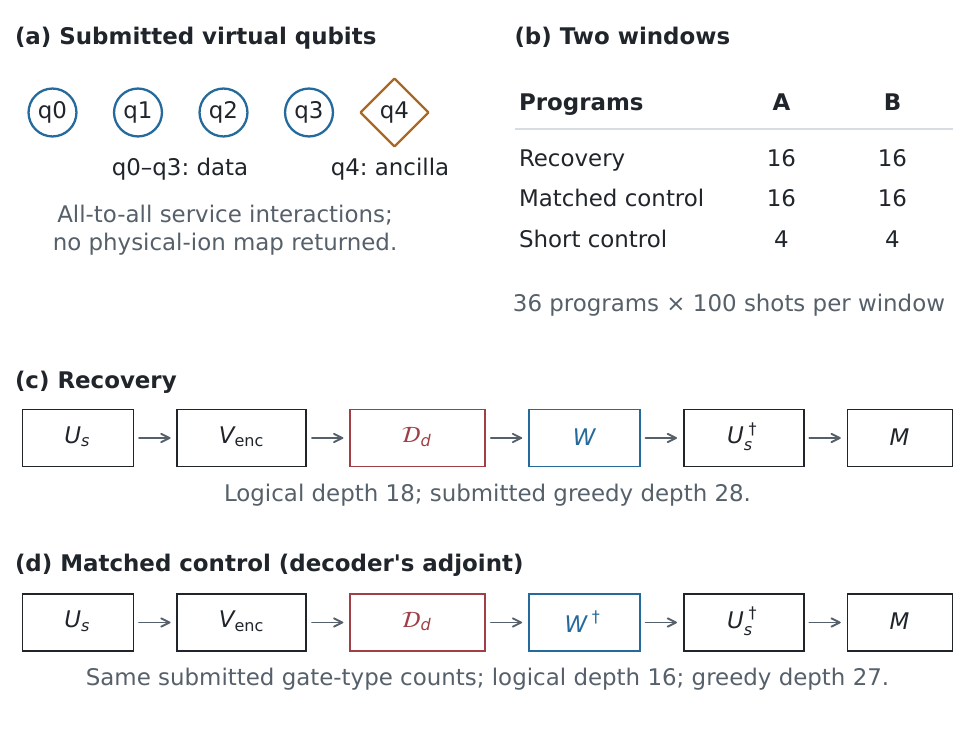}
  \caption{Recovery and controls on five submitted IonQ virtual qubits.
  Panel (a) shows the logical wire roles and advertised all-to-all
  interactions. Physical assignment remains unreported.
  Panel (b) lists the same 16 recovery circuits, 16 matched controls, and four short
  controls in each window. Window B reverses the submitted cell order.
  The lower panels compare the logical sequences: $V_{\mathrm{enc}}$ is
  the encoder, $\mathcal D_d$ is subsystem discard, and $W$ and
  $W^\dagger$ are the decoder and its adjoint. Here $M$ denotes
  computational-basis measurement of the selected output. Recovery and matched controls
  match submitted gate-type counts, with the depth differences shown.
  Subsystem discard is implemented by marginalization. Virtual
  qubit 4 is a zero-initialized logical ancilla for recovery and matched controls, but
  idle declared padding for the short controls.}
  \label{fig:ionq-panel}
\end{figure}

All 36 barrier-free OpenQASM~3 programs declared five virtual qubits and one
terminal measured bit, with 100 shots per program. The logical operations,
parameters and metadata agree cell by cell across windows; the QASM texts
differ only in a non-executable program-index comment. Window A used four
counterbalanced nine-program blocks, each containing four recovery--control
pairs covering all state and deletion labels plus one short control. Window B
reversed that submitted list (\cref{fig:ionq-panel}). The provider's chronological execution order remains unknown.
The provider returned aggregate binary probability histograms. Exact integer
counts were reconstructed for every program under prespecified numerical-integrity checks;
\cref{app:ionq-source-bindings} gives those checks and the circuit-identity
definition. 

Before either IonQ submission, we fixed the cellwise logical-circuit inventory,
its circuit-identity definition, both submitted cell orders, and the analysis plan for both windows.
We retrieved window-A outcomes before submitting window B; the two provider
tasks spanned 9--12 and 12--13 August 2026.  The second execution was therefore sequential and unblinded, with circuits,
order, and analysis already fixed.
The analysis plan was recorded locally, without public preregistration or external timestamping, and is held by the authors; the published
dataset contains the submitted sources and the cell tables, which include each
cell's circuit digest \citep{essayag2026reproducibility}.

For window $w\in\{A,B\}$, state $s\in\{0,1,2,3\}$, and deletion
$d\in\{0,1,2,3\}$, let
$\widehat p^{R}_{w,s,d}(0)$ and
$\widehat p^{\mathrm{adj}}_{w,s,d}(0)$ be the reconstructed
zero-outcome proportions for recovery and its matched control.
Let $k_{R,w}$ and $k_{\mathrm{adj},w}$ be the numbers of zero outcomes among the
1,600 recovery shots and the 1,600 control shots of window $w$.  The
equally weighted within-window estimators are
\begin{align}
  \widehat R_{\mathrm{IonQ}}^{(w)}
    &\coloneqq\frac1{16}\sum_{s=0}^{3}\sum_{d=0}^{3}\widehat p^{R}_{w,s,d}(0)
      =\frac{k_{R,w}}{1600},\\
  \widehat R_{\mathrm{adj}}^{(w)}
    &\coloneqq\frac1{16}\sum_{s=0}^{3}\sum_{d=0}^{3}
      \widehat p^{\mathrm{adj}}_{w,s,d}(0)
      =\frac{k_{\mathrm{adj},w}}{1600},\\
  \widehat\Delta_{\mathrm{adj}}^{(w)}
    &\coloneqq\widehat R_{\mathrm{IonQ}}^{(w)}-
      \widehat R_{\mathrm{adj}}^{(w)}
      =\frac{k_{R,w}-k_{\mathrm{adj},w}}{1600}.
  \label{eq:ionq-estimators}
\end{align}
Every cell has 100 shots, so equal cell weighting and pooling the 1,600 binary
outcomes agree numerically; counts and decisions remain window-specific.

Before either submission, we fixed the same two co-primary
criteria for each window.  Writing $k_R$ and $k_{\mathrm{adj}}$ for the corresponding
window's totals, their Hoeffding lower bounds were
\begin{align}
  L_R &\coloneqq \frac{k_R}{1600}-\sqrt{\frac{\log 40}{3200}},
  \label{eq:ionq-recovery-bound}\\
  L_{R-\mathrm{adj}} &\coloneqq \frac{k_R-k_{\mathrm{adj}}}{1600}
    -\sqrt{\frac{\log 40}{1600}}.
  \label{eq:ionq-control-bound}
\end{align}
Each bound applies Hoeffding's
inequality directly to independent bounded shot outcomes \citep{hoeffding1963probability}.  Each uses a one-sided significance level of $0.025$, so the union bound
gives simultaneous coverage of at least $95\%$ for the pair.  The decisions
were reported separately: recovery passed the prespecified two-thirds operational
reference if $L_R>2/3$, and the comparison of recovery with the matched control passed if
$L_{R-\mathrm{adj}}>0$.  Each window was described as clearing both criteria only when
both inequalities held.  The $95\%$ simultaneous
guarantee applies to each window's two inequalities; across all four inequalities in
the two-window conjunction, the union bound guarantees simultaneous coverage of at least $90\%$
under the stated assumptions.

\FloatBarrier
\subsection{Recovery passes both prespecified tests in two windows}
\label{sec:ionq-results}

All 36 programs completed in each IonQ window.  In window A, the 16 recovery
cells returned 1,457 successes in 1,600 shots; in window B they returned
1,476 successes.  Evaluating the same prespecified bound separately in each window
gave
\begin{align}
  \widehat R_{\mathrm{IonQ}}^{(A)}&=0.910625,
  &L_R^{(A)}&=0.876672,\nonumber\\
  \widehat R_{\mathrm{IonQ}}^{(B)}&=0.922500,
  &L_R^{(B)}&=0.888547.
  \label{eq:ionq-recovery-result}
\end{align}
Both lower bounds exceeded the prespecified two-thirds operational reference,
so the first co-primary criterion passed in each window.

The 16 matched controls returned 743 and 768 successes,
respectively, giving means of $0.464375$ and $0.480000$.  The two
recovery-minus-control estimates and their separately evaluated lower bounds
were
\begin{align}
  \widehat\Delta_{\mathrm{adj}}^{(A)}&=0.446250,
  &L_{R-\mathrm{adj}}^{(A)}&=0.398234,\nonumber\\
  \widehat\Delta_{\mathrm{adj}}^{(B)}&=0.442500,
  &L_{R-\mathrm{adj}}^{(B)}&=0.394484.
  \label{eq:ionq-control-result}
\end{align}
The second co-primary criterion also passed in both windows.
The required excess of 77 successes out of 1,600 is the statistical
threshold that the Hoeffding rule sets for detecting any positive
recovery-minus-control difference; the plan specified no minimum or expected effect size.  All 16
cellwise recovery-minus-control differences were positive in each window:
they ranged from $0.26$ to $0.66$ in A and from $0.29$ to $0.60$ in B.
For comparison, the ideal cellwise separations are $0.432237$ for states 0 and 3 and $0.621447$ for states 1 and 2, with mean $0.526842$
(\cref{sec:ionq-design}).
Descriptively, recovery means by designated deletion position ranged from $0.89$ to
$0.9225$ in A and from $0.905$ to $0.9325$ in B, showing high return across all four deletion positions.  The four short
controls returned 400 successes in 400 shots in each
window; because these controls are shorter and unpaired by deletion position,
their values are descriptive only.  Because each short control composes the
encoder with its exact inverse and provider-native compilation is unreported,
these scores do not show which operations were executed.
The prespecified decisions are summarized in \cref{tab:ionq-endpoints}.

\begin{table}[H]
  \centering
  \caption{Prespecified within-window endpoints for IonQ Forte Enterprise 1.
  Bounds are one-sided Hoeffding bounds at level $97.5\%$ and are evaluated
  separately in each window.}
  \label{tab:ionq-endpoints}
  \begin{tabular}{llrrc}
    \toprule
    Window & Endpoint & Estimate & Lower bound & Decision \\
    \midrule
    A & Recovery versus $2/3$ & 0.910625 & 0.876672 & Pass \\
    A & Recovery minus matched control & 0.446250 & 0.398234 & Pass \\
    B & Recovery versus $2/3$ & 0.922500 & 0.888547 & Pass \\
    B & Recovery minus matched control & 0.442500 & 0.394484 & Pass \\
    \bottomrule
  \end{tabular}
\end{table}

\begin{figure}[p]
  \centering
  \includegraphics[width=0.97\textwidth,
    alt={Paired IonQ plots and difference matrices show recovery above its control using the decoder's adjoint in all 16 cells of both windows. The caption explains the encodings and scope.}]{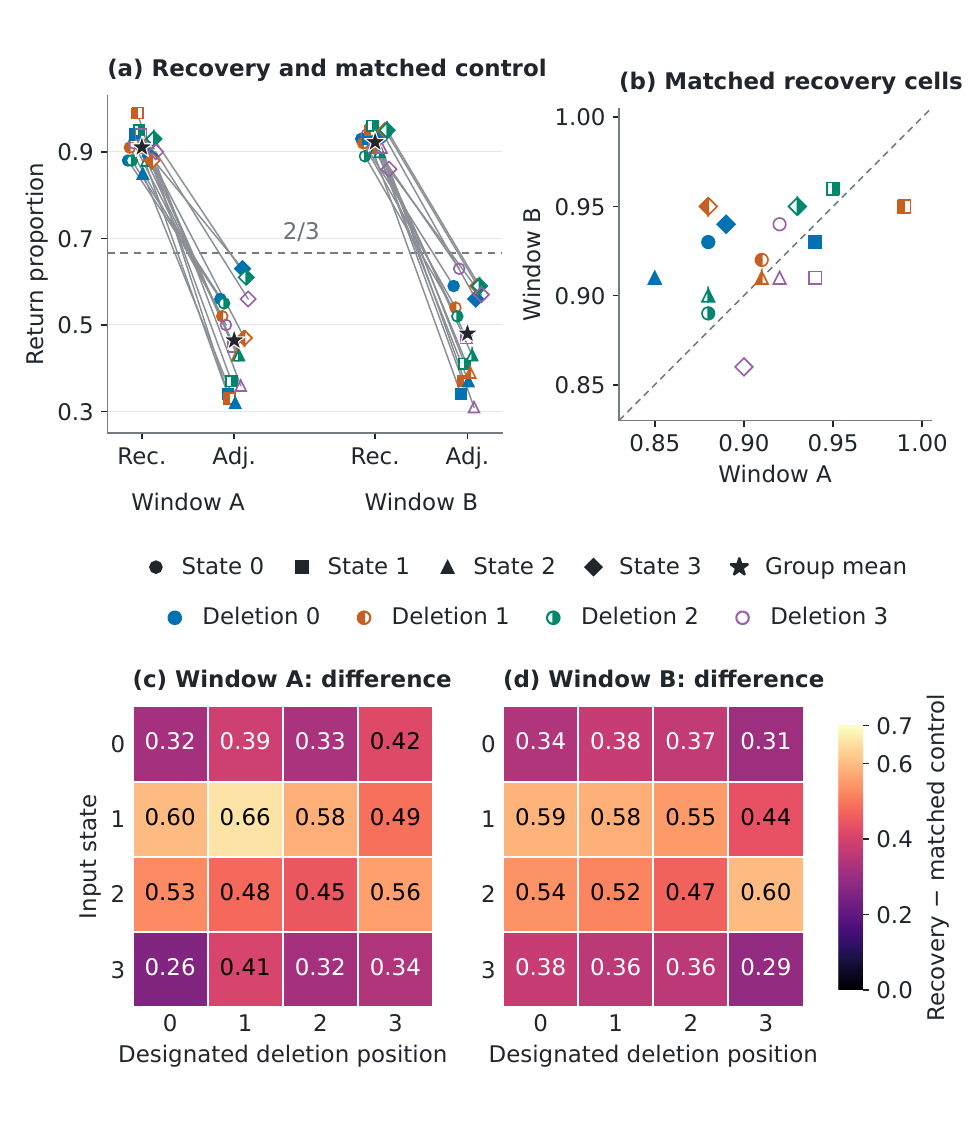}
  \caption{IonQ recovery exceeds the matched control in both windows.
  Panel (a) pairs 16 recovery--control cells per window: ``Rec.'' denotes recovery and ``Adj.'' the
  control using the decoder's adjoint, each gray segment joins the two circuits of one $(s,d)$ cell,
  small deterministic horizontal offsets separate categorical cells, and vertical coordinates are exact
  observed proportions; black stars mark
  four 16-cell means and the dashed line marks $2/3$.
  Panel (b) pairs recovery across windows, with a dashed equality diagonal.
  In both marker panels, shape identifies the
  input state, while deletion positions 0--3 use color and, respectively, full,
  left-half, right-half, and open fill; each non-mean marker represents a fixed
  100-shot circuit.
  Panels (c,d) give recovery minus the matched control on a common
  $0$--$0.70$ scale; each heat-map cell is the raw difference of its matched
  recovery and control circuits.  These displays show descriptive point estimates;
  \cref{tab:ionq-endpoints} reports the prespecified decisions and their one-sided
  bounds.}
  \label{fig:ionq-two-window}
\end{figure}

The cellwise and aggregate two-window comparisons are displayed in
\cref{fig:ionq-two-window}.

The recovery change from A to B, paired by the 16 state--deletion cells, was
$0.011875$.  Ten cell proportions increased, five decreased, and one was
unchanged; the cellwise changes ranged from $-0.04$ to $0.07$.  The
contrast (recovery minus the matched control) changed by $-0.003750$, and both short-control
means remained one.  Thus window B reproduced both prespecified
decisions and closely reproduced the aggregate contrast under a mirrored
submitted order.  The plan tested repetition of the two decisions; agreement between the scores in the two windows is descriptive.

\section{Discussion}
\label{sec:discussion}
\label{sec:architecture-synthesis}

\subsection{Workload and execution dependence}

The repeated experiments separate a score at one execution from the
question of whether it persists across executions. On IQM, descriptive route-level estimate
changes of opposite sign partly canceled in the aggregate RTSE mean, while the selected
communication effects failed their replication rules. The broad recovery
decline occurred later, under a different workload; its prespecified conditional shot-noise interval excludes zero, and its two executions also differed in submitted order. These observations show why a benchmark score needs its workload and
execution record. The discovery observations remain measurements of their
original executions, while the reruns constrain their repeatability.

Protocol benchmarks use concrete tasks to characterize useful processor
regions \citep{meirom2025protocols,mayo2026ibm,mayo2026crossplatform}.
Like the route-resolved do-nothing study of \citet{mayo2026howfar}, our length experiment varies swap distance; it compares that
protocol with RTSE through their common marginal at the initial qubit. Both showed a replicated
length drop, whereas the chosen directional test did not support better
retention by RTSE; descriptively, RTSE's score responded more strongly to route length. The predecessor's joint all-zero event deteriorated more
than its marginal: a shot with outcome zero at the initial qubit but a nonzero route
ancilla counts as a success for the marginal and as a failure for the all-zero event. The measured length
response therefore depends on the terminal event used to define success.

These results complement capability benchmarking's emphasis on circuit
structure and the full compilation and hardware stack
\citep{proctor2022capabilities,hines2024fullstack,proctor2025benchmarking}.
The communication comparisons fix the path and corresponding input or basis label, while resources,
output subsystems, and observables can differ. Where swapping scored above Bell-state transfer, as in the point estimates of both
executions, the difference ranks the two complete workloads; attribution to a physical mechanism requires a controlled
comparison. RTSE is a simple initial
diagnostic; communication and recovery tasks still need their own tests.

Quantum-error-correction circuits also serve as processor benchmarks
\citep{knill2001fivequbitbenchmark,wootton2020qecbenchmark}; recent work develops
detector-likelihood summaries for that purpose
\citep{hesner2025detectorlikelihood}. In our recovery experiment, the return
score includes preparation, encoding, compilation, routing where exposed,
discard, decoding, and readout. An exact code may have an encoder and decoder
too costly for a device, and a favorable average can be driven by one particularly
well-preserved input or noise direction while hiding poorer ones. Resolving all four inputs and
deletion positions, with placement and resource records, helps expose these
differences.

On IonQ, correct recovery scored above the deliberately incorrect matched control, which uses the decoder's adjoint, in every state--deletion cell of both windows (cellwise point estimates). The separation occurs with equal submitted gate-type counts; its ideal value for this circuit pair is $0.527$. The comparison
measures the difference between these two complete submitted sequences.

The scope of repetition also depends on the interface. IQM communication,
recovery and length experiments repeat byte-identical native sources at fixed physical locations; the recovery replication and length window B changed the submitted order.
IonQ repeats the logical-circuit inventory and decision rules under a reversed
submitted order. Its logical circuits agree cell by cell, while its OpenQASM texts differ only in a comment
that records list position (\cref{sec:ionq-design,app:ionq-source-bindings}).
For this submitted virtual-qubit diagnostic, both within-window decisions
recurred.

\subsection{Limitations}

The study covers one superconducting processor and one trapped-ion processor
under different experimental conditions.  IQM uses fixed physical routes and
placements; IonQ exposes only virtual qubits.  The device contrast combines architecture with differences in mapping,
routing, compilation, controls, and execution time.

The eight IQM communication routes and four recovery placements are fixed
diagnostic regions. Their intervals are conditional on those choices;
processor-wide prevalence requires a probability sample of regions. The length
result is likewise restricted to two vertex-disjoint geodesic route families
in two windows; estimating a general distance law requires broader sampling.

Each repeated experiment has two temporal observations. Estimating a drift
distribution or calibration half-life requires further independently scheduled
executions \citep{proctor2020drift}. Longitudinal zero-noise-extrapolation
experiments also show that measured effect sizes depend on execution time
\citep{koester2026claim}. Recovery took place after communication,
and the equal-window length analysis conditions on its two specific windows.
The IonQ rerun reproduces the within-window decisions; a numerical equivalence
claim would require a specified margin. The retained IQM capability-document times date the records,
while the underlying calibration ages remain unknown because per-calibration
timestamps were unavailable.

Submitted order also differed between executions: IQM recovery changed from
placement blocks to a balanced order, while IonQ and the length study reversed
their first-window lists.  The provider's chronological execution order remains unobserved. The
intervals quantify conditional shot noise and fixed-block heterogeneity under
the stated independence assumptions; temporal dependence is outside these
models.
IQM recovery-discovery intervals remain post hoc because the original recovery plan specified
only point estimands. The four IQM placements support descriptive comparisons
among those fixed placements. Deletion-position averages also combine different routing burdens and physical output
and readout carriers, which can contribute to their differences.

Parking and marginalizing a carrier implements a partial trace in ideal
circuit semantics. Physical loss, leakage, and disturbances caused by qubit
removal require corresponding hardware experiments.  The controls have further limits:
IQM's circuits that apply the encoder followed by its inverse are shorter than recovery, and IonQ's matched controls match gate-type counts while differing in operation order, rotation direction and
logical depth, with duration, native resources, physical placement, and readout interactions also
unmatched.  Establishing an improvement in state recovery from encoding, causal error suppression, or fault tolerance
requires additional baselines and tests.

RTSE measures return to the initial qubit. The separate one-way tests of the length study passed the
prespecified endpoint-sensitivity threshold, which indicates input-dependent endpoint signal under trusted preparation and
readout. RTSE and these tests have different resources and characterize their
respective circuits; whether the state arrives at the far endpoint in the RTSE circuit remains a
separate validation task. RTSE and the predecessor also differ in terminal
measurements even when compared through their common marginal at the initial qubit: the
predecessor measures every returned prefix ancilla.  Readout crosstalk from those extra
measurements can affect even the predecessor's marginal, and its all-zero score includes every ancilla.

\subsection{Next experiments}

The immediate priority is to estimate temporal distributions from several
independently scheduled executions; both recovery designs should now be repeated in this way, with the overall recovery score, the
minimum over deletion positions, and the temporal model fixed before data
collection.  A family of controls without recovery, matched in depth, duration, and
two-qubit gates, would help separate generic compiled-circuit burden from
recovery-specific structure.  Provider-side timing or finer partitioning into provider tasks would connect changes within
a session to actual execution time.

A controlled architecture comparison requires a common analysis plan and
comparable controls.  It must account for IQM's explicit sparse-topology
routing and IonQ's unobserved native resources and physical mapping.
Reporting native resources, measurement semantics, mapping visibility, and
statistical decisions alongside return scores would make these differences
assessable.

Finally, the length study should be extended from two geodesic route families
to a broader prospectively fixed route sample, with endpoint-sensitive controls
at more than one length and a model that accounts for native resources and duration.  The present result establishes a replicated length drop on two routes;
broader data would permit testing chip-wide behavior and length-decay models,
such as exponential fits.  RTSE can also be tested for predictive value: a
future design should select regions using RTSE data alone, specify performance
bands for teleportation and recovery, and then test whether those bands cover
the subsequent teleportation and recovery results.

\section{Conclusion}
\label{sec:conclusion}

Repeated executions of fixed IQM programs gave close aggregate RTSE scores
while individual route estimates moved in both directions. The estimated difference between RTSE and
teleportation shrank, while that between swapping and Bell-state transfer changed little. The
simultaneous confidence intervals for both differences included zero on repetition, and
all four communication targets failed their stringent replication rules.
RTSE thus supplies
an initial diagnostic whose average needs to be read alongside the
individual routes and the outcomes of the tasks being assessed.

The supporting length experiment shows that the RTSE score responds to route
length: extending two routes from 2 to 10 edges reduced both RTSE and
predecessor scores at the initial qubit in both windows. The measured
decrease was larger for RTSE. The predecessor's
score fell more when success was required for the whole route than when it
was required only at the initial qubit. Its measured length response therefore
depends on which outputs must succeed. Under trusted preparation and readout, separate one-way
tests detected endpoint signal that depends on the input; whether the state arrives at the far endpoint in the RTSE circuit itself remains open.

Recovery provided a direct test of an existing encoder and decoder after
subsystem discard. The IQM recovery mean declined on a repetition that also changed the
submitted order, and failed its prespecified replication test. On IonQ, recovery exceeded the two-thirds operational
reference and the control using the decoder's adjoint in each of two windows. Those
decisions concern the repeated logical circuits on submitted virtual qubits.
The control comparison tests sensitivity to the chosen decoder sequence; assessing an improvement in state recovery from encoding
requires an unencoded baseline, and recovery from physical qubit loss
requires a corresponding loss experiment.

The results characterize complete workloads at two executions per
experiment. Their broader use calls for independently scheduled repetitions,
larger route samples, and controls that account for depth, duration, and
native resources. A prospective design that uses RTSE alone to select regions
and then tests communication or recovery would assess the predictive
value of RTSE. Comparing architectures would additionally require common controls
and comparable information about compilation and physical mapping.

\appendix
\section{Conditional intervals and sensitivity analyses}
\label{app:statistical-details}

The main text defines the scientific endpoints and decision rules. This
appendix specifies how their conditional intervals are calculated. All
resampling conditions on the reported cells and windows. The binomial and
multinomial procedures use the stated models with independent shots; the
nonparametric procedures resample fixed blocks. These calculations describe
conditional shot noise or block heterogeneity; estimation of within-execution
drift requires additional timing or shot-level information.

\subsection{Communication endpoints and contrasts}
\label{app:communication-intervals}

Conditional on the 100 fixed program cells, simultaneous 95\% bounds were
obtained from 100,000 parametric binomial bootstrap replicates, using the
Jeffreys-smoothed cell proportion $(k+1/2)/(n+1)$ as the plug-in probability
\citep{jeffreys1946invariant,efron1979bootstrap}.  Here $k$ counts output-zero
successes for RTSE and teleportation, or $+1$ Pauli-product outcomes for
Bell-state transfer and frame-corrected $+1$ outcomes for swapping.
Write $\widetilde p_i$ for
this probability in cell $i$, and $w_{ji}$ for cell $i$'s weight in estimand
$j$.  Resamples use independent binomial counts at these fixed probabilities.
Each standardized error is centered at
$\widetilde\theta_j\coloneqq\sum_iw_{ji}\widetilde p_i$ and divided by the fixed
standard error $s_j$, where
$s_j^2\coloneqq\sum_iw_{ji}^2\widetilde p_i(1-\widetilde p_i)/n_i$ and $n_i$ is
the cell's shot count.  The route critical value uses the maximum signed
standardized error over its 28 members; the contrast critical value uses
the maximum absolute standardized error over its 18 members.  Each is the
95,000th ordered value from the 100,000 replicates.  Bounds use the raw
estimate minus the critical value times $s_j$; contrast upper bounds use
plus instead.  Each family has its own nominal 95\% level.  Pointwise intervals for averages within each execution
use each estimand's own absolute standardized-error quantile.

Bell-state transfer and swapping decisions also used a Bonferroni--Wilson bound
\citep{wilson1927probable}.  Within each overlap score, we assigned
$\alpha/3$ to each Pauli component, with $\alpha\coloneqq 0.05$, and computed
two-sided Wilson intervals for its positive-sign probability.  After
conversion to correlators by $2p-1$, the overlap lower bound combines the
lower $X$, upper $Y$, and lower $Z$ bounds in \cref{eq:bell-fidelity}.
The decision uses the smaller of this bound and the simultaneous bootstrap
lower bound.  The Wilson adjustment covers the three Pauli components within one score.
The simultaneous bootstrap handles route-level multiplicity.

\subsection{Joint length-study resampling}
\label{app:length-intervals}

The 13-member confirmatory family used 100,000 parametric max-statistic replicates
\citep{efron1979bootstrap}, with fixed Jeffreys-smoothed cell probabilities
\citep{jeffreys1946invariant}.  For a binary cell with $k$ successes in $n$
shots, we set $\widetilde p\coloneqq(k+1/2)/(n+1)$ and sampled
$K^*\sim\operatorname{Bin}(n,\widetilde p)$.  For a predecessor cell, the
three mutually exclusive categories were all-zero, outcome zero at the initial qubit but not
all-zero, and outcome one at the initial qubit.  Their observed counts $k_1,k_2,k_3$ gave
$\widetilde p_i\coloneqq(k_i+1/2)/(n+3/2)$, from which we sampled a multinomial
count vector with $n$ trials.  Each replicate draws counts at these fixed smoothed probabilities.

Let $\theta^*_{bj}$ be family member $j$ in replicate $b$,
$\widetilde\theta_j$ its value at the smoothed cell probabilities, and
$s_j$ the sample standard deviation of its replicate values (denominator
$100{,}000-1$).  In each replicate we computed
$M_b\coloneqq\max_{1\leq j\leq13}(\theta^*_{bj}-\widetilde\theta_j)/s_j$.
The critical value $c$ was the 95,000th ordered value of the 100,000
maxima.  The lower bound for member $j$ was
$\widehat\theta_j-cs_j$, where $\widehat\theta_j$ uses the raw observed
counts.

Within one multiplicity family, each exact window--cell pair is sampled once per
replicate and that draw is reused by every family member that contains it.
Thus the A, B, and equal-window members are evaluated jointly: an equal-window
replicate uses the same A and B draws at half weight, while distinct
window--cells remain conditionally independent.  The $L=0$
contrasts are outside the confirmatory family.  In their separate secondary
family, each window's set of eight shared $L=0$ cells is likewise sampled once
and reused across the three $L=0$ loss contrasts; different multiplicity
families use independently generated draws.  The eight-member exploratory adjacent-length family is constructed in the same way from its own draws.

\subsection{IQM recovery intervals}
\label{app:recovery-intervals}

The recovery-discovery plan specified point summaries only (\cref{sec:iqm-recovery}); their
subsequently added uncertainty intervals are labeled post hoc.  The
two-sided 95\% discovery block interval is a percentile interval from 100,000
nonparametric bootstrap replicates of the 16 fixed placement-by-state blocks,
preserving the four deletion positions within each selected block and using
linear $0.025$ and $0.975$ sample quantiles \citep{efron1979bootstrap}.  It describes block variation within the fixed placements under the
conditional model.
The descriptive pooled control interval is a shot-only 95\% Wilson
interval \citep{wilson1927probable}.
It treats the 16 control cells as a pooled shot sample and serves as a
descriptive summary alongside recovery.

The replication used the prespecified decision rule in
\cref{eq:qec-hoeffding}.  Its single primary decision pools the same 64
recovery cells with equal shot counts.  Secondary summaries include the 64
matched replication-minus-discovery cell changes, their mean absolute shift
($0.12313$) and root mean square shift ($0.14416$), the recovery-minus-control gap, and the fixed
placement and deletion-position summaries.  A conditional
binomial bootstrap uses 100,000 replicates, resampling 100 binary shots
independently within each fixed cell in each execution, using that
cell's raw observed success proportion as its binomial probability, and
recomputing the matched mean change. For the change in the recovery-minus-control
gap, each replicate also resamples the 16 control cells of each execution in the
same way, together with the 64 recovery cells of each execution, and recomputes the
difference between the two equal-cell recovery-minus-control means. Each reported
interval is the two-sided 95\% percentile interval using linear $0.025$ and $0.975$
sample quantiles.  A
separate two-sided 95\% interval for sensitivity to block heterogeneity uses 100,000 paired
nonparametric replicates of the 16 fixed placement-by-state block changes, retaining
all four deletion positions within each selected block and using the same
quantile convention \citep{efron1979bootstrap}.

\section{Experimental provenance and compilation records}
\label{app:experimental-provenance}

The following records support the fixed-workload comparisons: device names and
compilation records, the route-selection rule fixed before outcomes, and the source and
result checks. They supplement the four experiments in the main text.

\subsection{Devices, compilation, and provider-task chronology}
\label{app:execution-provenance}

The IQM device was \IQM{}; its six
executions were submitted as Amazon Braket OpenQASM 3 program-set provider tasks in
region \texttt{eu-north-1} \citep{aws2025iqmemerald}.  The IonQ device was IonQ Forte Enterprise 1; each
of its two windows was submitted in region \texttt{us-east-1} as
one provider task containing 36 ordered OpenQASM 3 programs.  The provider-task chronology is
given in \cref{tab:task-provenance}.  The authors retain the
corresponding provider records, which are not included in the public dataset.
Earlier Rigetti work tested the submission procedure and is excluded from
scientific inference.

\begin{table}[H]
  \centering
  \caption{Provider-task chronology.  Times are UTC; every requested shot was
  returned successfully.  ``Created--ended'' is the provider task-lifecycle
  span between its recorded timestamps; it may include queueing and service
  bookkeeping as well as device execution.}
  \label{tab:task-provenance}
  {\small\begin{tabular}{lcl}
    \toprule
    Provider task & Programs $\times$ shots & Created--ended \\
    \midrule
    Communication discovery & $100\times200$
      & Aug. 2 16:51:59--Aug. 3 00:00:36 \\
    Communication replication & $100\times200$
      & Aug. 4 10:45:22--10:45:43 \\
    Recovery discovery & $80\times100$
      & Aug. 6 20:26:10--20:26:45 \\
    Recovery replication & $80\times100$
      & Aug. 7 14:44:00--14:44:31 \\
    IonQ recovery window A & $36\times100$
      & Aug. 9 18:58:48--Aug. 12 00:31:16 \\
    IonQ recovery window B & $36\times100$
      & Aug. 12 17:27:15--Aug. 13 14:04:04 \\
    RTSE length window A & $96\times200$
      & Aug. 14 12:44:00--12:44:37 \\
    RTSE length window B & $96\times200$
      & Aug. 14 15:57:29--15:58:03 \\
    \bottomrule
  \end{tabular}
  }
\end{table}

The IQM circuits were routed and compiled offline by a deterministic compiler.  The published
dataset \citep{essayag2026reproducibility} contains the exact
submitted QASM sources and derived tables; private provider records and the
intermediate logical and routed sources are excluded from it and are held by the authors.  The submitted sources use
physical qubit indices inside full-verbatim OpenQASM 3 boxes and contain only
native \texttt{PRX} and \texttt{CZ} gates before terminal measurement; the requested qubit mapping was fixed throughout submission.

A complete IQM device-capability document was retained before each IQM submission.
The provider properties exposed calibration values but no per-calibration
timestamps, so the age of the underlying calibration cannot be certified.
For the recovery replication, the live check verified complete value coverage
and topological eligibility on the four already fixed paths.

\subsection{Initial route selection and repair}
\label{app:route-selection}

The four initial qubits were fixed spatial landmarks from IQM's published numbered
Crystal-54 geometry \citep{aws2025iqmemerald}: peripheral initial qubits 39 and 15, and central initial qubits 19 and 35.
Pre-submission checks on the live device retained these spatial landmark choices.  Let $G_{\mathrm{initial}}$ be
the complete undirected coupling graph captured on 28 July 2026 for the
original route selection.  All 54 active qubits and 85 edges in that IQM snapshot were eligible.  The diagram supplied the
spatial landmarks; this live graph supplied the eligible routes.

For each initial qubit, we enumerated every simple six-edge geodesic.  Let $H$ be the
eligible graph used at the current selection step.  For two candidate
routes $\pi,\pi'$ with endpoints $u,u'$, let
$E(\pi)$ and $V(\pi)$ denote, respectively, the undirected edge set and vertex
set of route $\pi$, and let
\begin{equation}
  d_{\mathrm{end}}\coloneqq d_H(u,u'),\qquad
  d_E\coloneqq\lvert E(\pi)\mathbin{\triangle}E(\pi')\rvert,
  \label{eq:route-selection-metrics}
\end{equation}
and let $\nu_H(u,u')$ be the number of shortest $u$--$u'$ paths.  Candidate
pairs were ranked lexicographically by decreasing $d_{\mathrm{end}}$,
decreasing $d_E$, increasing $\nu_H(u,u')$, and then the lexicographically
smallest ordered route pair.  This rule used $H\coloneqq G_{\mathrm{initial}}$ at the
original selection, where the two routes of each selected pair were labeled $j=0$ and $j=1$ in lexicographic order; at the repair below, each replacement kept the label of the route it replaced, so the final labels need not follow lexicographic order.

Two old routes through absent qubit 25 later became
ineligible.  The repair step used the 53-node graph $G_{\mathrm{live}}$ of
\cref{sec:communication-design}. With
$H\coloneqq G_{\mathrm{live}}$, the three graph criteria were evaluated against the
surviving sibling route before any lexicographic tie-break. The replacement for initial qubit 15 was the unique topology optimum. For replacements from initial qubit 19 tied
in those three criteria, the
repaired route pairs from initial qubits 15 and 19 were first required to
share no undirected edge; shared nodes were allowed.  The remaining candidates
maximized
\begin{equation}
  Q(\pi)\coloneqq\sum_{e\in E(\pi)}\log f_{\mathrm{CZ}}(e)
         +\sum_{v\in V(\pi)}\log f_{\mathrm{RO}}(v),
  \label{eq:route-calibration-score}
\end{equation}
with a lexicographically smallest-route tie-break.  Here $f_{\mathrm{CZ}}$ and
$f_{\mathrm{RO}}$ are the live provider fields \texttt{fCZ} and \texttt{fRO}.
These fields came from the capability document captured at
07:58:39~UTC on 2 August, before the communication submission and before any
protocol outcome.  The document's service update time was
06:10:21~UTC, but no per-metric timestamp was exposed, so metric age is
unknown. The pairs from initial qubits 35 and 39 retained the
routes originally fixed from topology.  No circuit outcome entered either selection.
The final routes are listed in \cref{tab:routes} and drawn on the captured
graph in \cref{fig:frozen-routes}.

\subsection{Source identity in the length study}
\label{app:length-source-bindings}

Each returned result carried a provider format header that the submitted
programs lacked; the embedded source remained OpenQASM 3.  A validation
amendment, fixed after the window-A results returned and
before any window-A outcome was inspected or window B was submitted, allowed
only this header to be removed before checking source equality; the raw
provider records were preserved.  The same rule applied to both windows.  
\subsection{IonQ circuit identity and count reconstruction}
\label{app:ionq-source-bindings}

Each window submitted the same 36-cell logical-circuit
inventory in one provider task with 100 shots per program (\cref{sec:ionq-design}).  For each cell,
the circuit identity is a SHA-256 digest of a canonical description containing the virtual-qubit
count, the ordered operations with their qubits and parameters, and the circuit
metadata, but not the position in the submitted list; this digest
agrees cell by cell across windows.  The matched OpenQASM texts
differ only in a non-executable comment that records each program's position in the submitted list,
which window B mirrors; the executable operations agree exactly.  The provider returned
aggregate measurement-probability histograms; per-shot records were
unavailable.  For every program, the prespecified rule
reconstructed exact integer counts from the returned binary probabilities.
Each probability times 100 had to lie within $10^{-9}$ of an integer, the two
probabilities had to sum to one within $10^{-12}$, and the reconstructed counts
had to sum exactly to 100.  We applied no mitigation or postselection; the retained records do not show whether the provider processed the outcomes before returning the histograms.

\subsection{Excluded spatial grouping}
\label{app:excluded-spatial-summary}

A metadata error labeled initial qubits $\{15,19\}$ as boundary and $\{35,39\}$ as
interior; the published geometry instead classifies $\{15,39\}$ as peripheral
and $\{19,35\}$ as central.  We exclude all summaries based on the erroneous
grouping, including its prespecified replication contrast and descriptive
interval.  Initial qubit and path selection, compilation, both submitted orders, and all primary
and multiplicity-controlled endpoints are unchanged. The
numerical results resolved by placement, state, and deletion position in the tables and
figures are unchanged. The exclusion covers the grouping comparison itself,
including a post-hoc substitute based on the corrected labels.

\subsection{Reproducibility dataset}
\label{app:reproducibility-dataset}

A reproducibility dataset is publicly available on Zenodo at
\href{https://doi.org/10.5281/zenodo.21969397}{doi:10.5281/zenodo.21969397}
(version 0.2.0) \citep{essayag2026reproducibility}; it was deposited under this article's title.  It contains
the submitted QASM sources for all 624 program executions, the aggregate data and the derived cell tables
that provide every plotted value and aggregate, the figures, and the analysis, figure-generation and
validation code for the 101,600 requested shots.  Data, figures,
and QASM sources are released under CC BY 4.0 and code under the MIT License.  Raw provider
result objects, private provider records and account records are not included.  This manuscript updates presentation while retaining the
archived numerical evidence and submitted QASM inventory.

Reproducibility corrections for version 0.2.0 accompany this arXiv version
as ancillary material. They preserve the archived observations, submitted
QASM, and scientific results; the Zenodo DOI above identifies the original
dataset.

\begingroup
\renewcommand{\bibsection}{\section*{\refname}\addcontentsline{toc}{section}{\refname}}
\emergencystretch=3em
\bibliographystyle{unsrtnat}
\bibliography{references}

\begin{thebibliography}{44}
\providecommand{\natexlab}[1]{#1}
\providecommand{\url}[1]{\texttt{#1}}
\expandafter\ifx\csname urlstyle\endcsname\relax
  \providecommand{\doi}[1]{doi: #1}\else
  \providecommand{\doi}{doi: \begingroup \urlstyle{rm}\Url}\fi

\bibitem[Cross et~al.(2019)Cross, Bishop, Sheldon, Nation, and
  Gambetta]{cross2019quantumvolume}
Andrew~W. Cross, Lev~S. Bishop, Sarah Sheldon, Paul~D. Nation, and Jay~M.
  Gambetta.
\newblock Validating quantum computers using randomized model circuits.
\newblock \emph{Physical Review A}, 100\penalty0 (3):\penalty0 032328, 2019.
\newblock \doi{10.1103/PhysRevA.100.032328}.

\bibitem[Blume-Kohout and Young(2020)]{blumekohout2020volumetric}
Robin Blume-Kohout and Kevin~C. Young.
\newblock A volumetric framework for quantum computer benchmarks.
\newblock \emph{Quantum}, 4:\penalty0 362, 2020.
\newblock \doi{10.22331/q-2020-11-15-362}.

\bibitem[Proctor et~al.(2022{\natexlab{a}})Proctor, Rudinger, Young, Nielsen,
  and Blume-Kohout]{proctor2022capabilities}
Timothy Proctor, Kenneth Rudinger, Kevin Young, Erik Nielsen, and Robin
  Blume-Kohout.
\newblock Measuring the capabilities of quantum computers.
\newblock \emph{Nature Physics}, 18\penalty0 (1):\penalty0 75--79,
  2022{\natexlab{a}}.
\newblock \doi{10.1038/s41567-021-01409-7}.

\bibitem[Proctor et~al.(2025)Proctor, Young, Baczewski, and
  Blume-Kohout]{proctor2025benchmarking}
Timothy Proctor, Kevin Young, Andrew~D. Baczewski, and Robin Blume-Kohout.
\newblock Benchmarking quantum computers.
\newblock \emph{Nature Reviews Physics}, 7\penalty0 (2):\penalty0 105--118,
  2025.
\newblock \doi{10.1038/s42254-024-00796-z}.

\bibitem[Lubinski et~al.(2023)Lubinski, Johri, Varosy, Coleman, Zhao, Necaise,
  Baldwin, Mayer, and Proctor]{lubinski2023application}
Thomas Lubinski, Sonika Johri, Paul Varosy, Jeremiah Coleman, Luning Zhao,
  Jason Necaise, Charles~H. Baldwin, Karl Mayer, and Timothy Proctor.
\newblock Application-oriented performance benchmarks for quantum computing.
\newblock \emph{IEEE Transactions on Quantum Engineering}, 4:\penalty0 1--32,
  2023.
\newblock \doi{10.1109/TQE.2023.3253761}.

\bibitem[Hines and Proctor(2024)]{hines2024fullstack}
Jordan Hines and Timothy Proctor.
\newblock Scalable full-stack benchmarks for quantum computers.
\newblock \emph{IEEE Transactions on Quantum Engineering}, 5:\penalty0 1--12,
  2024.
\newblock \doi{10.1109/TQE.2024.3404502}.

\bibitem[Murali et~al.(2019)Murali, Baker, Javadi-Abhari, Chong, and
  Martonosi]{murali2019noiseadaptive}
Prakash Murali, Jonathan~M. Baker, Ali Javadi-Abhari, Frederic~T. Chong, and
  Margaret Martonosi.
\newblock Noise-adaptive compiler mappings for noisy intermediate-scale quantum
  computers.
\newblock In \emph{Proceedings of the Twenty-Fourth International Conference on
  Architectural Support for Programming Languages and Operating Systems}, pages
  1015--1029, 2019.
\newblock \doi{10.1145/3297858.3304075}.

\bibitem[Li et~al.(2025)Li, Zhou, and Feng]{li2025qknob}
Sanjiang Li, Xiangzhen Zhou, and Yuan Feng.
\newblock Benchmarking quantum circuit transformation with {QKNOB} circuits.
\newblock \emph{IEEE Transactions on Quantum Engineering}, 6:\penalty0 1--15,
  2025.
\newblock \doi{10.1109/TQE.2025.3527399}.

\bibitem[Dasgupta and Humble(2022)]{dasgupta2022reproducibility}
Samudra Dasgupta and Travis~S. Humble.
\newblock Characterizing the reproducibility of noisy quantum circuits.
\newblock \emph{Entropy}, 24\penalty0 (2):\penalty0 244, 2022.
\newblock \doi{10.3390/e24020244}.

\bibitem[Huang et~al.(2021)Huang, Lin, Ku, and Chen]{huang2021statetransfer}
Yi-Te Huang, Jhen-Dong Lin, Huan-Yu Ku, and Yueh-Nan Chen.
\newblock Benchmarking quantum state transfer on quantum devices.
\newblock \emph{Physical Review Research}, 3:\penalty0 023038, 2021.
\newblock \doi{10.1103/PhysRevResearch.3.023038}.

\bibitem[Zhukov et~al.(2019)Zhukov, Kiktenko, Elistratov, Pogosov, and
  Lozovik]{zhukov2019protocolbenchmarks}
A.~A. Zhukov, E.~O. Kiktenko, A.~A. Elistratov, W.~V. Pogosov, and Yu.~E.
  Lozovik.
\newblock Quantum communication protocols as a benchmark for programmable
  quantum computers.
\newblock \emph{Quantum Information Processing}, 18\penalty0 (1):\penalty0 31,
  2019.
\newblock \doi{10.1007/s11128-018-2144-y}.
\newblock Published online 6 December 2018.

\bibitem[Meirom et~al.(2025)Meirom, Mor, and Weinstein]{meirom2025protocols}
Dekel Meirom, Tal Mor, and Yossi Weinstein.
\newblock Benchmarking quantum computers via protocols.
\newblock Preprint,
  \href{https://arxiv.org/abs/2505.12441v2}{arXiv:2505.12441v2} [quant-ph],
  2025.

\bibitem[Mayo et~al.(2026{\natexlab{a}})Mayo, Mor, and Weinstein]{mayo2026ibm}
Nitay Mayo, Tal Mor, and Yossi Weinstein.
\newblock Benchmarking quantum computers via protocols, comparing {IBM}'s
  {Heron} vs {IBM}'s {Eagle}.
\newblock Preprint,
  \href{https://arxiv.org/abs/2603.04377v3}{arXiv:2603.04377v3} [quant-ph],
  2026{\natexlab{a}}.

\bibitem[Mayo et~al.(2026{\natexlab{b}})Mayo, Mor, and
  Weinstein]{mayo2026crossplatform}
Nitay Mayo, Tal Mor, and Yossi Weinstein.
\newblock Benchmarking quantum computers via protocols, comparing
  superconducting and ion-trap quantum technology.
\newblock Preprint,
  \href{https://arxiv.org/abs/2603.27397v3}{arXiv:2603.27397v3} [quant-ph],
  2026{\natexlab{b}}.

\bibitem[M{\'a}rquez et~al.(2025)M{\'a}rquez, Sierra-Sosa, and
  Garc{\'e}s]{marquez2025teleportation}
Cristian M{\'a}rquez, Daniel Sierra-Sosa, and Kelly Garc{\'e}s.
\newblock A teleportation protocol variant for single-{QPU} benchmarking.
\newblock \emph{IEEE Access}, 13:\penalty0 209266--209281, 2025.
\newblock \doi{10.1109/ACCESS.2025.3639914}.

\bibitem[Mayo et~al.(2026{\natexlab{c}})Mayo, Mor, and {Zabokritskiy
  (Yohananov)}]{mayo2026howfar}
Nitay Mayo, Tal Mor, and Aryeh~Lev {Zabokritskiy (Yohananov)}.
\newblock How far can you do nothing on a quantum computer?
\newblock Preprint,
  \href{https://arxiv.org/abs/2608.21904v2}{arXiv:2608.21904v2} [quant-ph],
  2026{\natexlab{c}}.

\bibitem[Peters et~al.(2022)Peters, Shyamsundar, Li, and
  Perdue]{peters2022timereversal}
Evan Peters, Prasanth Shyamsundar, Andy C.~Y. Li, and Gabriel Perdue.
\newblock Qubit assignment using time reversal.
\newblock \emph{PRX Quantum}, 3\penalty0 (4):\penalty0 040333, 2022.
\newblock \doi{10.1103/PRXQuantum.3.040333}.

\bibitem[Proctor et~al.(2022{\natexlab{b}})Proctor, Seritan, Rudinger, Nielsen,
  Blume-Kohout, and Young]{proctor2022mirror}
Timothy Proctor, Stefan Seritan, Kenneth Rudinger, Erik Nielsen, Robin
  Blume-Kohout, and Kevin Young.
\newblock Scalable randomized benchmarking of quantum computers using mirror
  circuits.
\newblock \emph{Physical Review Letters}, 129\penalty0 (15):\penalty0 150502,
  2022{\natexlab{b}}.
\newblock \doi{10.1103/PhysRevLett.129.150502}.

\bibitem[Throckmorton and Das~Sarma(2020)]{throckmorton2020swap}
Robert~E. Throckmorton and S.~Das~Sarma.
\newblock Fidelity of a sequence of {SWAP} operations on a spin chain.
\newblock \emph{Physical Review B}, 102:\penalty0 035439, 2020.
\newblock \doi{10.1103/PhysRevB.102.035439}.
\newblock Erratum: Physical Review B 104, 119902 (2021),
  \href{https://doi.org/10.1103/PhysRevB.104.119902}{doi:10.1103/PhysRevB.104.119902}.

\bibitem[Miguel-Ramiro et~al.(2026)Miguel-Ramiro, Illiano, Mazza, Pirker,
  Freund, Cacciapuoti, Caleffi, and D{\"u}r]{miguelramiro2026qping}
Jorge Miguel-Ramiro, Jessica Illiano, Francesco Mazza, Alexander Pirker, Julia
  Freund, Angela~Sara Cacciapuoti, Marcello Caleffi, and Wolfgang D{\"u}r.
\newblock {QPing}: A quantum ping primitive for quantum networks.
\newblock \emph{IEEE Journal on Selected Areas in Communications}, 44:\penalty0
  4997--5011, 2026.
\newblock \doi{10.1109/JSAC.2026.3693981}.

\bibitem[McKay et~al.(2023)McKay, Hincks, Pritchett, Carroll, Govia, and
  Merkel]{mckay2023layerfidelity}
David~C. McKay, Ian Hincks, Emily~J. Pritchett, Malcolm Carroll, Luke C.~G.
  Govia, and Seth~T. Merkel.
\newblock Benchmarking quantum processor performance at scale.
\newblock Preprint,
  \href{https://arxiv.org/abs/2311.05933v1}{arXiv:2311.05933v1} [quant-ph],
  2023.

\bibitem[Lozano~Palacio et~al.(2026)Lozano~Palacio, Nayfeh, Ware, and
  McKay]{lozano2026layerfidelity}
Maria~Jose Lozano~Palacio, Hasan Nayfeh, Matthew Ware, and David~C. McKay.
\newblock Parameter analysis and optimization of layer fidelity for quantum
  processor benchmarking at scale.
\newblock \emph{IEEE Transactions on Quantum Engineering}, 7:\penalty0 1--10,
  2026.
\newblock \doi{10.1109/TQE.2026.3668098}.

\bibitem[Hagiwara and Nakayama(2020)]{hagiwara2020deletion}
Manabu Hagiwara and Ayumu Nakayama.
\newblock A four-qubits code that is a quantum deletion error-correcting code
  with the optimal length.
\newblock In \emph{2020 IEEE International Symposium on Information Theory
  (ISIT)}, pages 1870--1874, 2020.
\newblock \doi{10.1109/ISIT44484.2020.9174339}.

\bibitem[Nakayama and Hagiwara(2020)]{nakayama2020singledeletion}
Ayumu Nakayama and Manabu Hagiwara.
\newblock Single quantum deletion error-correcting codes.
\newblock In \emph{2020 International Symposium on Information Theory and Its
  Applications (ISITA)}, volume~65 of \emph{IEICE Proceeding Series}, pages
  329--333, 2020.
\newblock \doi{10.34385/proc.65.B10-1}.
\newblock Preprint version: arXiv:2004.00814v1.

\bibitem[Massar and Popescu(1995)]{massar1995optimal}
Serge Massar and Sandu Popescu.
\newblock Optimal extraction of information from finite quantum ensembles.
\newblock \emph{Physical Review Letters}, 74\penalty0 (8):\penalty0 1259--1263,
  1995.
\newblock \doi{10.1103/PhysRevLett.74.1259}.

\bibitem[Hothem et~al.(2024)Hothem, Young, Catanach, and
  Proctor]{hothem2024capability}
Daniel Hothem, Kevin Young, Tommie Catanach, and Timothy Proctor.
\newblock Learning a quantum computer's capability.
\newblock \emph{IEEE Transactions on Quantum Engineering}, 5:\penalty0 1--26,
  2024.
\newblock \doi{10.1109/TQE.2024.3430215}.

\bibitem[Renes et~al.(2004)Renes, Blume-Kohout, Scott, and Caves]{renes2004sic}
Joseph~M. Renes, Robin Blume-Kohout, A.~J. Scott, and Carlton~M. Caves.
\newblock Symmetric informationally complete quantum measurements.
\newblock \emph{Journal of Mathematical Physics}, 45\penalty0 (6):\penalty0
  2171--2180, 2004.
\newblock \doi{10.1063/1.1737053}.

\bibitem[Scott(2006)]{scott2006tight}
A.~J. Scott.
\newblock Tight informationally complete quantum measurements.
\newblock \emph{Journal of Physics A: Mathematical and General}, 39\penalty0
  (43):\penalty0 13507--13530, 2006.
\newblock \doi{10.1088/0305-4470/39/43/009}.

\bibitem[Horodecki et~al.(1999)Horodecki, Horodecki, and
  Horodecki]{horodecki1999teleportation}
Micha\l{} Horodecki, Pawe\l{} Horodecki, and Ryszard Horodecki.
\newblock General teleportation channel, singlet fraction, and
  quasidistillation.
\newblock \emph{Physical Review A}, 60\penalty0 (3):\penalty0 1888--1898, 1999.
\newblock \doi{10.1103/PhysRevA.60.1888}.

\bibitem[Nielsen(2002)]{nielsen2002average}
Michael~A. Nielsen.
\newblock A simple formula for the average gate fidelity of a quantum dynamical
  operation.
\newblock \emph{Physics Letters A}, 303\penalty0 (4):\penalty0 249--252, 2002.
\newblock \doi{10.1016/S0375-9601(02)01272-0}.

\bibitem[{\.Z}ukowski et~al.(1993){\.Z}ukowski, Zeilinger, Horne, and
  Ekert]{zukowski1993eventready}
Marek {\.Z}ukowski, Anton Zeilinger, Michael~A. Horne, and Artur~K. Ekert.
\newblock {``Event-ready-detectors''} {Bell} experiment via entanglement
  swapping.
\newblock \emph{Physical Review Letters}, 71\penalty0 (26):\penalty0
  4287--4290, 1993.
\newblock \doi{10.1103/PhysRevLett.71.4287}.

\bibitem[Bennett et~al.(1993)Bennett, Brassard, Cr{\'e}peau, Jozsa, Peres, and
  Wootters]{bennett1993teleportation}
Charles~H. Bennett, Gilles Brassard, Claude Cr{\'e}peau, Richard Jozsa, Asher
  Peres, and William~K. Wootters.
\newblock Teleporting an unknown quantum state via dual classical and
  {Einstein-Podolsky-Rosen} channels.
\newblock \emph{Physical Review Letters}, 70\penalty0 (13):\penalty0
  1895--1899, 1993.
\newblock \doi{10.1103/PhysRevLett.70.1895}.

\bibitem[{Amazon Web Services}(2026)]{awsbraketionq}
{Amazon Web Services}.
\newblock {IonQ} trapped ion quantum computing.
\newblock Amazon Braket Quantum Computers, 2026.
\newblock URL \url{https://aws.amazon.com/braket/quantum-computers/ionq/}.
\newblock Accessed 26 August 2026.

\bibitem[Essayag and {Zabokritskiy
  (Yohananov)}(2026)]{essayag2026reproducibility}
Isaac~Barouch Essayag and Aryeh~Lev {Zabokritskiy (Yohananov)}.
\newblock Reproducibility package for {From Round-Trip State Echo to Error
  Recovery: Snapshot-Resolved Quantum-Hardware Diagnostics}.
\newblock Dataset, Zenodo, version 0.2.0, 2026.
\newblock URL \url{https://doi.org/10.5281/zenodo.21969397}.

\bibitem[Mohammad et~al.(2025)Mohammad, Govindarajan, Komar, and
  Seegerer]{aws2025iqmemerald}
Zia Mohammad, Charunethran~Panchalam Govindarajan, Peter Komar, and Stefan
  Seegerer.
\newblock Amazon {Braket} launches new 54-qubit superconducting quantum
  processor from {IQM}.
\newblock AWS Quantum Technologies Blog, July 2025.
\newblock URL
  \url{https://aws.amazon.com/blogs/quantum-computing/amazon-braket-launches-new-54-qubit-superconducting-quantum-processor-from-iqm/}.
\newblock Accessed 16 August 2026.

\bibitem[Hoeffding(1963)]{hoeffding1963probability}
Wassily Hoeffding.
\newblock Probability inequalities for sums of bounded random variables.
\newblock \emph{Journal of the American Statistical Association}, 58\penalty0
  (301):\penalty0 13--30, 1963.
\newblock \doi{10.1080/01621459.1963.10500830}.

\bibitem[Knill et~al.(2001)Knill, Laflamme, Martinez, and
  Negrevergne]{knill2001fivequbitbenchmark}
E.~Knill, R.~Laflamme, R.~Martinez, and C.~Negrevergne.
\newblock Benchmarking quantum computers: The five-qubit error correcting code.
\newblock \emph{Physical Review Letters}, 86\penalty0 (25):\penalty0
  5811--5814, 2001.
\newblock \doi{10.1103/PhysRevLett.86.5811}.

\bibitem[Wootton(2020)]{wootton2020qecbenchmark}
James~R. Wootton.
\newblock Benchmarking near-term devices with quantum error correction.
\newblock \emph{Quantum Science and Technology}, 5\penalty0 (4):\penalty0
  044004, 2020.
\newblock \doi{10.1088/2058-9565/aba038}.

\bibitem[Hesner et~al.(2025)Hesner, Het{\'e}nyi, and
  Wootton]{hesner2025detectorlikelihood}
Ian Hesner, Bence Het{\'e}nyi, and James~R. Wootton.
\newblock Using detector likelihood for benchmarking quantum error correction.
\newblock \emph{Physical Review A}, 111\penalty0 (5):\penalty0 052452, 2025.
\newblock \doi{10.1103/PhysRevA.111.052452}.

\bibitem[Proctor et~al.(2020)Proctor, Revelle, Nielsen, Rudinger, Lobser,
  Maunz, Blume-Kohout, and Young]{proctor2020drift}
Timothy Proctor, Melissa Revelle, Erik Nielsen, Kenneth Rudinger, Daniel
  Lobser, Peter Maunz, Robin Blume-Kohout, and Kevin Young.
\newblock Detecting and tracking drift in quantum information processors.
\newblock \emph{Nature Communications}, 11\penalty0 (1):\penalty0 5396, 2020.
\newblock \doi{10.1038/s41467-020-19074-4}.

\bibitem[K{\"o}ster and Mauerer(2026)]{koester2026claim}
Dominik K{\"o}ster and Wolfgang Mauerer.
\newblock Claim against measurement: Statistical artefacts in quantum error
  mitigation benchmarks.
\newblock arXiv:2605.29872v2, 2026.

\bibitem[Jeffreys(1946)]{jeffreys1946invariant}
Harold Jeffreys.
\newblock An invariant form for the prior probability in estimation problems.
\newblock \emph{Proceedings of the Royal Society of London. Series A.
  Mathematical and Physical Sciences}, 186\penalty0 (1007):\penalty0 453--461,
  1946.
\newblock \doi{10.1098/rspa.1946.0056}.

\bibitem[Efron(1979)]{efron1979bootstrap}
Bradley Efron.
\newblock Bootstrap methods: Another look at the jackknife.
\newblock \emph{The Annals of Statistics}, 7\penalty0 (1):\penalty0 1--26,
  1979.
\newblock \doi{10.1214/aos/1176344552}.

\bibitem[Wilson(1927)]{wilson1927probable}
Edwin~B. Wilson.
\newblock Probable inference, the law of succession, and statistical inference.
\newblock \emph{Journal of the American Statistical Association}, 22\penalty0
  (158):\penalty0 209--212, 1927.
\newblock \doi{10.1080/01621459.1927.10502953}.

\end{thebibliography}
\endgroup

\end{document}